\documentclass[preprint]{aastex}
\usepackage{graphicx}
\usepackage{dcolumn}
\usepackage{bm}
\usepackage{epsfig}
\usepackage{graphicx}
\usepackage{hyperref}
\usepackage[usenames]{color}
\usepackage{url}
\usepackage{listings}
\usepackage{mathrsfs}

\hypersetup{
    colorlinks=true,
    linkcolor=red,
    citecolor=blue,
}

\newcommand{\remove}[1]{}

\def\ie{{\it i.e.}}
\def\eg{{\it e.g.}}

\def\be{\begin{equation}}
\def\ee{\end{equation}}
\def\ba{\begin{eqnarray}}
\def\ea{\end{eqnarray}}

\def\mgh{{\tt MGHalofit}}
\def\hf{{\tt Halofit}}

\newcommand{\loge}{{\rm ln~}}
\newcommand{\rmd}{{\rm d}}
\newcommand{\dln}{{\rm d~ln~}}
\newcommand{\dsqln}{{\rm d^2~ln~}}

\shortauthors{G. -B. Zhao}

\shorttitle{MGHalofit}

\begin{document}

\title{Modeling the nonlinear clustering in modified gravity models I: \\ A fitting formula for matter power spectrum of $f(R)$ gravity}
\author{Gong-Bo Zhao \altaffilmark{1,2}}

\email{gongbo@icosmology.info}

\altaffiltext{1}{National Astronomy Observatories,
Chinese Academy of Science, Beijing, 100012, P.R.China}

\altaffiltext{2}{Institute of Cosmology and Gravitation, University of Portsmouth,
Portsmouth, PO1 3FX, UK}

\begin{abstract} 

Based on a suite of $N$-body simulations of the Hu-Sawicki model of $f(R)$ gravity with different sets of model and cosmological parameters, we develop a new fitting formula with a numeric code, \mgh, to calculate the nonlinear matter power spectrum $P(k)$ for the Hu-Sawicki model. We compare the \mgh~predictions at various redshifts ($z\leqslant1$) to the $f(R)$ simulations and {find} that the {relative error of the \mgh~fitting formula of $P(k)$ is no larger than }6\% at $k\leqslant1$ h/Mpc and 12\% at $k\in(1,10]$ h/Mpc respectively. Based on a sensitivity study of an ongoing and a future spectroscopic survey, we estimate the detectability of a signal of modified gravity described by the Hu-Sawicki model using the power spectrum up to quasi-nonlinear scales. \mgh~is publicly available at \url{http://icosmology.info/website/MGHalofit.html}.  

\end{abstract}


\maketitle

\section{Introduction}
\label{sec:intro}

Understanding the accelerating expansion of the universe, which was first discovered using the supernova measurements \citep{Riess,Perlmutter}, is one of the key problems in modern sciences. Since the cosmic acceleration challenges Einstein's theory of General Relativity (GR) without the cosmological constant, which predicts a decelerating universe, there have been much effort in modifying Einstein's theory by either adding a new Dark Energy (DE) component in the framework of GR (see \citealt{DEreview} and references therein for the DE review), or changing GR itself (for a thorough review of modified gravity, see \citealt{MGreview} and references therein). It is true that both approaches can give the same evolution of the universe at the background level, they generally predict different structure formation. Given the ongoing and upcoming cosmological large scale structure (LSS) surveys mapping the 3D structure growth in the universe, it is possible to break the theoretical degeneracy between DE and modified gravity (MG).  

{ In some MG models, \eg, the generalised symmetron, dilaton \citep{sim:symmetron+dilaton} and chameleon \citep{sim:chameleon} models, the linear perturbation theory fails to predict the matter power correctly even on linear scales. Therefore ignoring the nonlinearities or modeling the nonlinearities improperly may result in diluted, or even biased cosmological constraints.}       

The attempts to model the nonlinear clustering analytically goes back to \citet{HKLM} based on the {\it scaling Ansatz} of the clustering evolution, and it was later generalised and refined by \citet{PD94,JMW95,PD96}. However, the fitting formulae based on the scaling Ansatz were later found inaccurate when contrasted to simulations \citep{PD96fail1,PD96fail2,PD96fail3,PD96fail4}.

The {\it halo model} is an alternative approach to model the nonlinear clustering, in which the large scale clustering is calculated using the correlation between different halos while the small scale clustering is derived using the convolution of the dark matter profile with the halo itself (\citealt{halo1,halo2,halo3,halo4,halo5,halo6,halo7,halo8}; for a review of the halo model, see \citealt{haloreview}). A new fitting formula called \hf~was developed based on the halo model and calibrated using $N$-body simulations for cold dark matter (CDM) cosmologies \citep{halofit}.  \hf~has a much better accuracy than the previous fitting formulae and it was commonly used since developed. However, since
\hf~was calibrated using simulations for CDM models, in which the growth generally has no scale-dependence, the default \hf~is not applicable to models with a scale-dependent growth, \eg, models with massive neutrinos and most modified gravity models.  

\hf~has been extended for models with massive neutrinos \citep{halofitmnu}, but not yet for modified gravity models \footnote{For a different halo model approach to model $P(k)$ for the HS model, see \citet{Pkhalo}.}. Given that cosmological tests of gravity is one of the key science drivers for most of the upcoming LSS surveys, in this paper we shall make the first attempt to develop a nonlinear $P(k)$ fitting formula, \mgh, for one specific $f(R)$ model, the Hu-Sawicki model \citep{HS} (HS hereafter), which represents a class of MG models with the chameleon mechanism \citep{chameleon}. \mgh~is calibrated by high-resolution $N$-body simulations, and it is used to estimate the detectability of MG using LSS surveys.

This paper is structured as follows. In Sec. 2 the basics of the HS model are presented. Sec. 3 is devoted to the $N$-body simulations of the HS model for various model and cosmological parameters, followed by the development of {\tt MGHalofit} in Sec. 4. Discussions and summary are in the final section.

\section{The Hu-Sawicki model}
\label{sec:HS}

The Hu-Sawicki model \citep{HS} is one of the viable $f(R)$ models that is capable {of fitting} current observations after {tuning} the model parameters. On large scales, it mimics a $\Lambda$CDM background cosmology while on nonlinear scales it can evade the solar system tests by the naturally built-in chameleon mechanism \citep{chameleon}. On intermediate scales, it has distinctive observational features that can be tested using cosmological experiments (\citealt{HSobs1,HSobs2,HSobs9,PS,HSobs8,HSobs10,HSobs4,HSobs6,HSobs7,HSobs5,HSobs3}; see \citealt{HSobsrev} for a review).

The action of the model is, \be\label{eq:S} S=\int \sqrt{-g} \left [\frac{f(R)}{16\pi G} + \mathcal{L}_{\rm M} \right ]{\rm d}^4 x\ee   where \be\label{eq:fR} f(R)=-m^2\frac{\alpha_1(-R/m^2)^n}{\alpha_2(-R/m^2)^n+1},~~~~m^2=H_0^2\Omega_{\rm M}\ee

Variation of the action $S$ with respect to the metric yields the equation of motion for a new scalar degree of freedom $f_R\equiv\frac{\rmd f(R)}{\rmd R}$. As \citet{quasistatic} pointed out, the quasi-static approximation (QSA) is valid for this model which means that the time derivative of the scalar field can be ignored. Under the QSA, the equation of motion of the scalar field $\delta f_R$ can be obtained as, \be\label{eq:DfR} \nabla^2  \delta f_R = -\frac{a^2}{3}[\delta R(f_R)+8\pi G\delta\rho_{\rm M}]\ee where $\delta f_R=f_R(R)-f_R(\bar{R}), \delta R=R-\bar{R}$ and $\delta\rho_{\rm M} - \bar{\rho_{\rm M}}$. One can invert Eq. (\ref{eq:fR}) to relate $R$ to $f_R$, namely, \be\label{eq:f_R} f_R=-\frac{\alpha_1}{\alpha_2^2}\frac{n(-R/m^2)^{n-1}}{[(-R/m^2)^n+1]^2}\simeq -\frac{n\alpha_1}{\alpha_2^2}\left(\frac{m^2}{-R}\right)^{n+1}\ee  where the approximation holds if the background cosmology is close to a $\Lambda$CDM model, and in this case, one can approximate $\bar{R}$ as,
\be \bar{R}\simeq 3 H_0^2 \left[\Omega_{M}(1+z)^{3}+4\Omega_{\Lambda} \right]\ee At redshift $z=0$, 
\be\label{eq:R0bar} \bar{R}_0\equiv\bar{R}(z=0)\simeq 3 H_0^2 \left(1+3\Omega_{\Lambda} \right) \ee where a flat universe is assumed. 

Combining Eqs (\ref{eq:f_R}) and (\ref{eq:R0bar}), one can rewrite Eq (\ref{eq:f_R}) in terms of $f_{R0}$, which is the background value of $f_R$ at redshift $z=0$, as,
\be f_R\simeq f_{R0}\left[\frac{3H_0^2(1+\Omega_{\Lambda})}{-R}  \right]^{n+1}\ee and $\delta R$ is given explicitly as, \be\label{eq:DR}\delta R(f_R)=3H_0^2\left\{(1+3\Omega_{\Lambda})\left(\frac{f_{R0}}{f_R}\right)^{\frac{1}{n+1}} -\left[\Omega_M(1+z)^3+4\Omega_{\Lambda}\right] \right\}\ee The scalar field $f_R$ can then be solved numerically by combining Eqs (\ref{eq:DfR}) with (\ref{eq:DR}), given the model parameters $f_{R0}$ and $n$ with background cosmological parameters.

The modified Poisson equation for the gravitational potential $\Phi$ can be obtained by adding up the $00$ and $ii$ component of the modified Einstein equation in the HS model, namely, \be\label{eq:Poisson} \nabla^2\Phi=\frac{16\pi G}{3}a^2\delta\rho_{\rm M}+\frac{a^2}{6}\delta R(f_R)\ee The dynamics of the system is determined by Eqs. (\ref{eq:DfR}) and (\ref{eq:Poisson}). Eq (\ref{eq:DfR}) is a nonlinear Poisson equation and it has to be solved numerically on regular or self-adaptive grids using iteration methods \citep{Oyaizu1,Zhao11,ECOSMOG,MGGadget}. But qualitatively, we can see that this model has the following features  \citep{HS,Oyaizu2,Oyaizu3,Zhao11,ECOSMOG}, 
\begin{description}
\item[(A)] In the low density region (on large scales), where $|f_R|\sim|\bar{f}_R|$, \ie, $\delta f_R\sim0$, $\delta R(f_R)$ can be linearised and $f_R$ can be easily solved in Fourier space. In this case, gravity is locally enhanced by $1/3$ within the Compton wavelength; 
\item[(B)] In the high density region (on nonlinear scales), where $|f_R|\sim0$, GR is locally recovered. 
\end{description}

Since the strength of gravity (essentially the mass of the scalar field) varies with local density, this is called the chameleon mechanism, which is key to evade the {solar system tests}. 

\section{$N$-body simulations of the Hu-Sawicki model}

The nonlinear clustering in the HS model cannot be accurately calculated without performing large $N$-body simulations due to the complicated nonlinear dynamics of the chameleon mechanism in this model. 
The $N$-body simulations for the HS model were first performed by \citet{Oyaizu1,Oyaizu2} using a particle-mesh (PM) code with regular grids, based on which the halo statistics of this model was first analysed in \citet{Oyaizu3}. The same model was re-simulated and exploited by \citet{Zhao11} \footnote{Details of this simulation project and the visualisation including images and movies are publicly available at \url{http://icosmology.info/website/Nbody_Simulation.html}} using a modified version of {\tt MLAPM} \citep{MLAPM}, by \citet{HSPk} using {\tt ECOSMOG} \citep{ECOSMOG}, which is a variant of {\tt Ramses} \citep{Ramses}, and by \citet{MGGadget} using {\tt MGGADGET}, a modified version of the tree code {\tt GADGET} \citep{GADGET}. Thanks to the self-adaptive grid structure of {\tt MLAPM} and {\tt ECOSMOG} and to the tree structure in {\tt MGGADGET}, a much higher force resolution (up to a factor of $7$ improvement compared to \citealt{Oyaizu1,Oyaizu2}) was obtained in these new simulations. 

The previous HS simulations are based on the WMAP seven-year best fit cosmology \citep{WMAP7} (WMAP7 hereafter, parameters summarised in Eq (\ref{eq:wmap7})). In this work, we need new HS simulations for other background cosmologies to calibrate \mgh~to make it robust for a range of cosmological parameters. We choose to run new HS simulations using {\tt ECOSMOG} for the Planck \citep{Planck} (summarised in Eq (\ref{eq:planck})) and WMAP nine-year \citep{WMAP9} (WMAP9, summarised in Eq (\ref{eq:wmap9})) best fit cosmologies for the calibration because they sizably differ from the WMAP7 cosmology, \eg, $\Omega_{\rm M}^{\rm Planck}$ and  $\Omega_{\rm M}^{\rm WMAP9}$ is larger than  $\Omega_{\rm M}^{\rm WMAP7}$ by 28\% and 7\% respectively. We use the previous WMAP7 simulation \citep{Zhao11} as well for the calibration. 
 \ba  
 {\rm Planck:} &&\{\Omega_b, \Omega_c, \Omega_k, h, n_s, \sigma_8\} \label{eq:planck}     =\{0.04825,0.2589, 0,0.678,0.961,0.84\} \\
 {\rm WMAP9:} &&\{\Omega_b, \Omega_c, \Omega_k, h, n_s, \sigma_8\} \label{eq:wmap9}  =\{0.04363,0.2136, 0,0.718,0.973,0.80\} \\ 
 {\rm WMAP7:} &&\{\Omega_b, \Omega_c, \Omega_k, h, n_s, \sigma_8\} \label{eq:wmap7}  =\{0.04181,0.1982, 0,0.730,0.958,0.80\}
 \ea
 For each set of parameters, we simulate three $f(R)$ models with $n=1, |f_{R0}|=10^{-4,-5,-6}$ (F4, F5, F6 models hereafter). We simulate the $|f_{R0}|=0$ ($\Lambda$CDM) model as well using the same initial condition to make direct comparison. 

We generate the initial conditions (IC) at $z=49$ using {\tt MPgrafic} \citep{MPgrafic}, which is a parallel version of {\tt Grafic}, an IC generator in the {\tt COSMICS} package \citep{COSMICS}. We use the {\tt ECOSMOG} code to perform the simulation using $256^3$ particles in a box with $B=128$ Mpc/h a side \footnote{As tested in \citet{Zhao11}, both Hu-Sawicki and $\Lambda$CDM simulations using the box size B=128 Mpc/h agree with those using B=256 and 64 Mpc/h on scales $k\in[0.05,~10]$ h/Mpc. This demonstrates that the B=128 Mpc/h simulation result used in this work is robust.}. More parameters for the simulation are summarized in Table \ref{Tab:sim}. {The power spectrum of the simulation is measured using the {\tt POWMES} code \citep{POWMES}, whose precision is within sub-percent level on scales we are interested in.} For more technical details of the simulation and data analysis, we refer the readers to \citet{Zhao11,ECOSMOG}.

\begin{table}

\begin{center}
\begin{tabular}{ccccccc}

\hline\hline
Cosmology & log$_{10}|f_{R0}|$  &Seeds & $\sqrt[3]{\rm N_p}$ & Box [Mpc/h] & output $z$ & Reference \\ \hline
Planck & $-4,-5,-6,0$    & 1&256  &128 & 1, 0.8, 0.6, 0.4, 0.2, 0  & This work  \\
WMAP9 & $-4,-5,-6,0$  & 1&256  &128 & 1, 0.8, 0.6, 0.4, 0.2, 0  & This work  \\
WMAP7 & $-4,-5,-6,0$  & 10&256   &128 & 1, 0 & \citet{Zhao11} \\
\hline\hline

\end{tabular}
\end{center}
\caption{The details of the simulations used for the calibration where $N_p$ is the number of the particles.}
\label{Tab:sim}
\end{table}%

\begin{figure*}
  \begin{center}
  \includegraphics[scale=0.28]{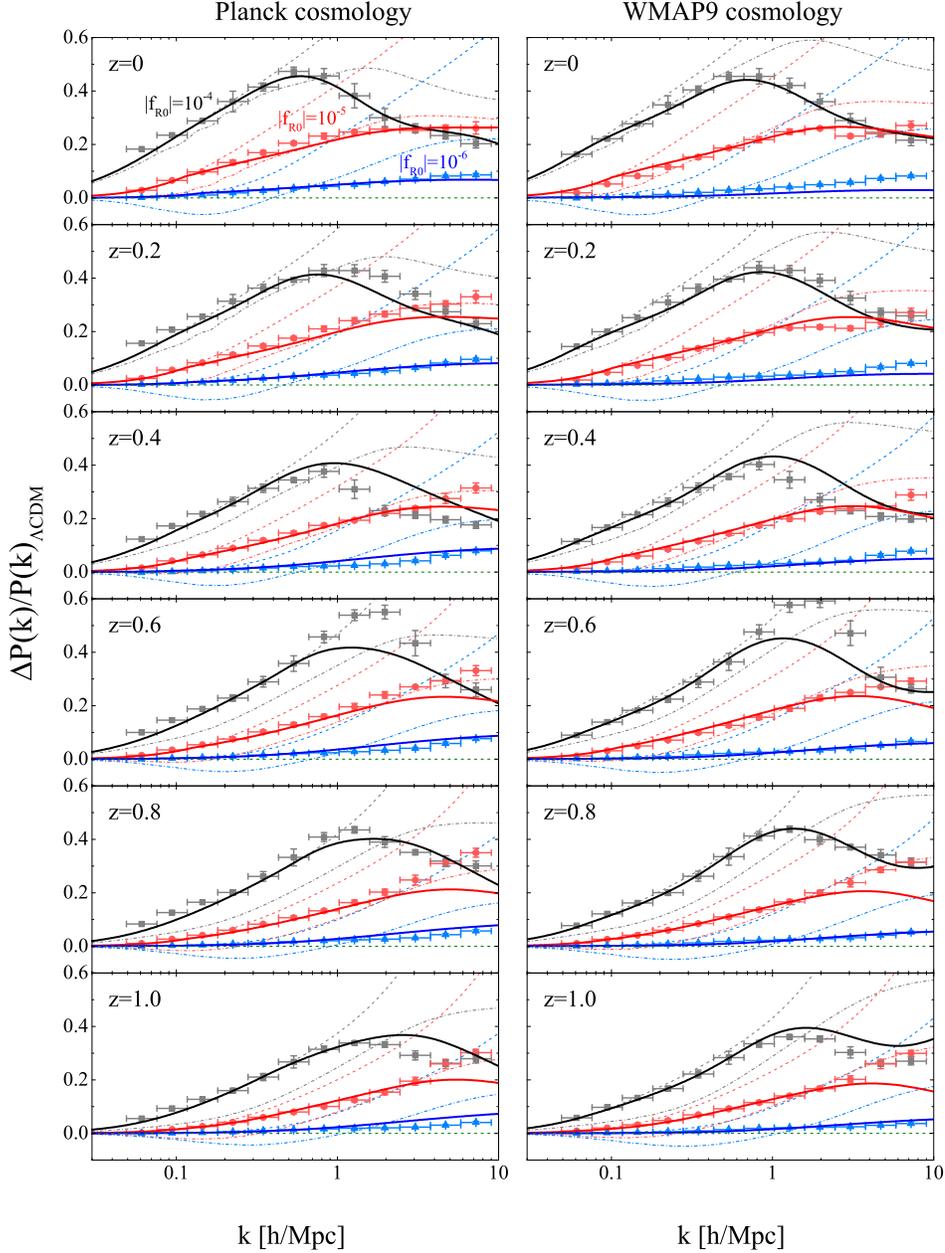}
  \end{center}
  \caption{The fractional difference in matter power spectrum between the Hu-Sawicki gravity model and the $\Lambda$CDM model, \ie, $\Delta_P(k)\equiv\Delta P(k)/P(k)_{\rm \Lambda CDM}$ in the Planck (left panels) and WMAP9 (right) cosmologies. From {top to bottom}, the panels show the result from redshifts $z=0$ to $z=1$ with a $\Delta z=0.2$ increment. In each panel, from {top to bottom}, the black, red and blue error bars and curves stand for the $f(R)$ model with log$_{10}|f_{R0}|=-4,-5$ and $-6$. {The data points with error bars show the $N$-body simulation result, and the curves are:} thick solid: \mgh; thin dashed: linear perturbation theory calculated using {\tt MGCAMB}; thin dash-dotted: Halofit prediction. The horizontal green dashed line illustrates $\Delta_P=0$ to guide eyes.} \label{fig:planck+wmap9}
\end{figure*}

\begin{figure*}
  \begin{center}
  \includegraphics[scale=0.25]{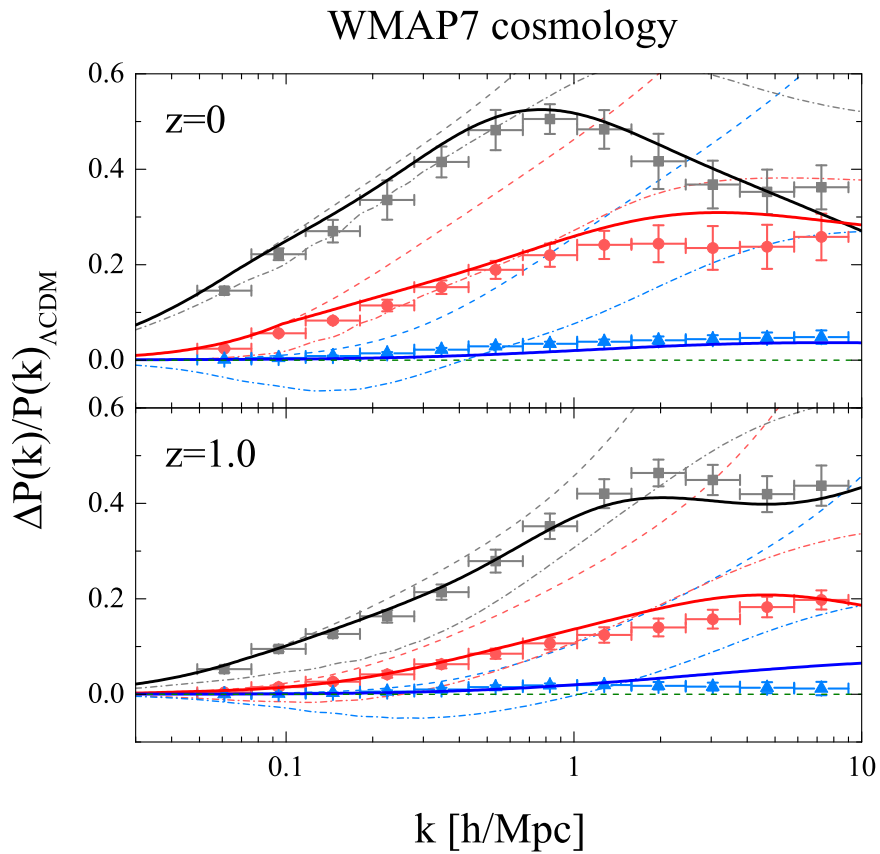}
  \end{center}
  \caption{Same as Fig \ref{fig:planck+wmap9} but for the WMAP7 cosmology. The upper and lower panels are for redshifts $z=0$ and $1$ respectively. } \label{fig:wmap7}
\end{figure*}

\begin{figure*}
  \begin{center}
  \includegraphics[scale=0.3]{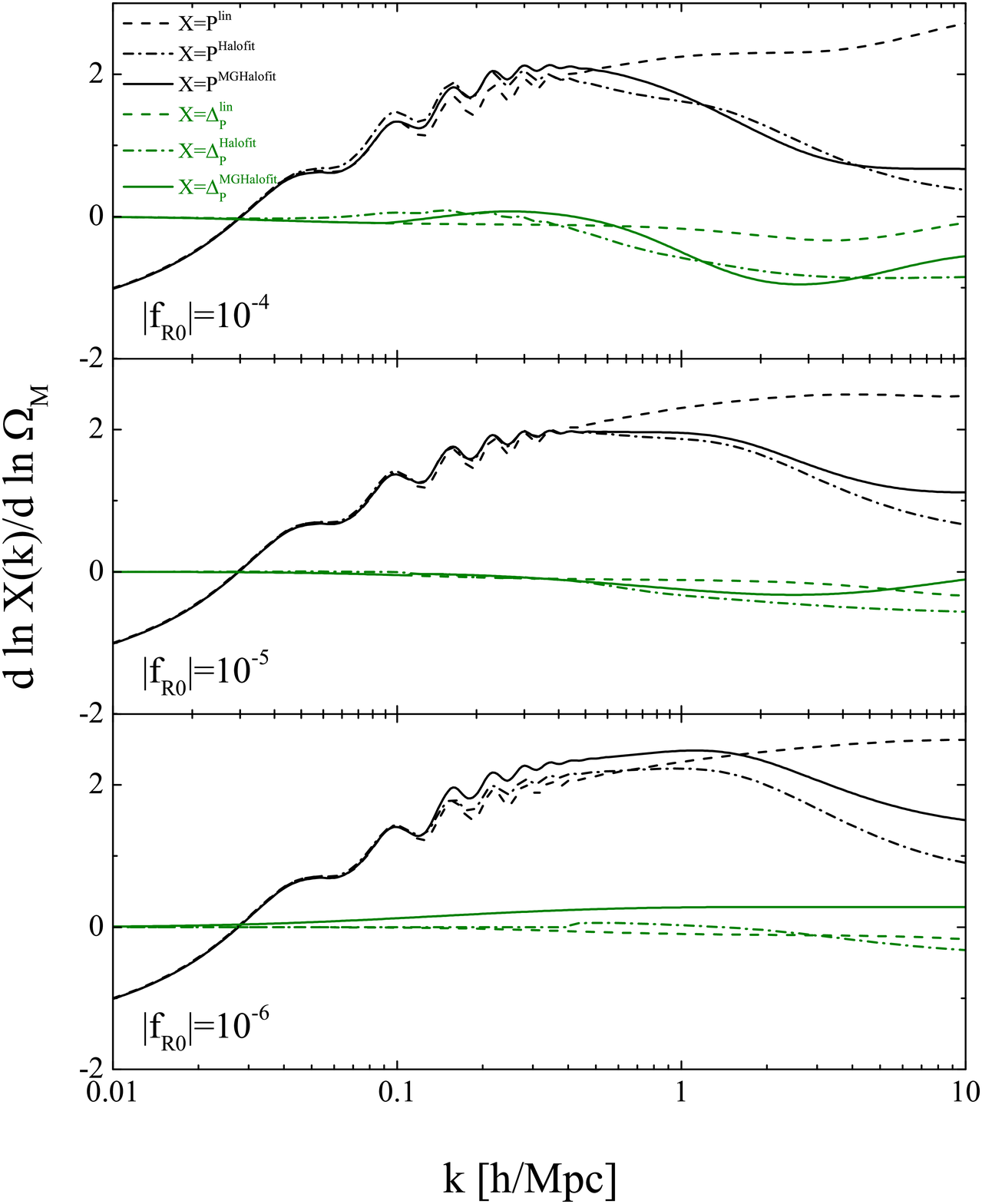}
  \end{center}
  \caption{The sensitivity of the (fractional) matter power spectrum $(\Delta_P)~P(k)$ to $\Omega_{\rm M}$ illustrated by the quantity (d ln$\Delta_P(k)$/d ln $\Omega_{\rm M}$) d ln$P(k)$/d ln $\Omega_{\rm M}$. The linear, \hf~and \mgh~predictions are shown in dashed, dash-dotted and solid curves respectively.} \label{fig:deriv}
\end{figure*}

The simulation results are shown in data points with error bars in Figs \ref{fig:planck+wmap9} and \ref{fig:wmap7}. We show the fractional difference \be\label{eq:DP} \Delta_P(k)\equiv \frac{P(k)_{\rm HS}}{P(k)_{\rm \Lambda{CDM}}}-1\ee instead of the power spectrum itself to eliminate the sample variance. As seen in Figs \ref{fig:planck+wmap9} and \ref{fig:wmap7}, the dependence of $\Delta_P(k)$ is much stronger on $|f_{R0}|$ than on other cosmological parameters, \eg, $\Omega_{\rm M}$. This is illustrated more clearly in Fig \ref{fig:deriv} where we show the quantities of dln$P(k)/$dln$\Omega_{\rm M}$ and dln$\Delta_P(k)/$dln$\Omega_{\rm M}$. This essential shows the response of the fractional change in $P(k)$ or $\Delta_P(k)$ with respect to the fractional change in $\Omega_{\rm M}$. As we can see, \be \left\vert\frac{{\rm d~ln}P(k)}{{\rm d~ln}\Omega_{\rm M}}\right\vert > \left\vert\frac{{\rm d~ln}\Delta_P(k)}{{\rm d~ln}\Omega_{\rm M}}\right\vert \ee in all cases. Due to the fact that $\Delta_P(k)$ is almost immune to the sample variance \citep{Oyaizu1,Oyaizu2,Oyaizu3,Zhao11,ECOSMOG} and its weak dependence on $\Omega_{\rm M}$, we choose to fit $\Delta_P(k)$ rather than $P(k)$ measured from simulations when calibrating the \mgh~parameters.

\section{A new fitting formula for the matter power spectra for the HS model}
\label{sec:fittingformula}

The {\tt Halofit} fitting formula for CDM cosmologies was developed by \citet{halofit} based on the halo model approach, and it was recently re-calibrated using simulations with a better resolution \citep{halofitnew}.  {\tt Halofit} has been extensively used to calculate the nonlinear power spectra for the $\Lambda$CDM-like cosmologies, in which the growth function is scale-independent. However, for models where the growth is scale-dependent, \eg, models with massive neutrinos ($m_{\nu}$CDM hereafter) and the modified gravity models especially for the HS model, there is a large discrepancy between the {\tt Halofit} prediction and the $N$-body simulation result \citep{Oyaizu2,Oyaizu3,Zhao11}.     

A fitting formula for the nonlinear power spectrum of the $m_{\nu}$CDM model has been developed by extending the {\tt Halofit} formula. Specifically, new parameters multiplying ${f_\nu}$, the fractional energy density for massive neutrinos, are added to the formula and $N$-body simulations are used for the calibration \citep{halofitmnu}. In this work, we adopt a similar approach for the HS model.  

\subsection{Generalisation of the functional form of {\tt Halofit}}

In \hf, the dimensionless power spectra $\Delta^2$ is split into the quasi-nonlinear ($\Delta^2_{\rm Q}$) and nonlinear terms ($\Delta^2_{\rm H}$), dominating the power on large and small scales respectively.  

\be\Delta^2\equiv\frac{k^3P(k)}{2\pi^2}=\Delta^2_{\rm Q}+\Delta^2_{\rm H}\ee
\ba\Delta^2_{\rm Q}(k)&=&\Delta^2_{\rm L}(k)\frac{[1+{\Delta}^2_{\rm L}(k)]^{{\beta}(n_{\rm eff},\mathcal{C})}}{1+{\alpha}(n_{\rm eff},\mathcal{C}){\Delta}^2_{\rm L}(k)}{\rm exp}\left[-\left(y/4+y^2/8\right)\right] \nonumber\\
{\Delta}^2_{\rm H}(k)&=&\frac{{\Delta}^{2'}_{\rm H}(k)}{1+{\mu}(n_{\rm eff},\mathcal{C})/y+{\nu}(n_{\rm eff},\mathcal{C})/y^2}\nonumber\\
{\Delta}^{2'}_{\rm H}(k)&=&\frac{{a}(n_{\rm eff},\mathcal{C})y^{3f_1(\Omega_{\rm M})}}{1+{b}(n_{\rm eff},\mathcal{C})y^{f_2(\Omega_{\rm M})}
+[{c}(n_{\rm eff},\mathcal{C})f_3(\Omega_{\rm M})y]^{3-{\gamma}(n_{\rm eff},\mathcal{C})}}\ea 

where $\Delta_{\rm L}$ is the dimensionless linear power spectrum and, \ba\label{eq:shapepara} \sigma^2(R,z)&=&\int \Delta_{\rm L}^2(k,z) {\rm exp}(-k^2R^2)\dln{k} \nonumber \\
n_{\rm eff}&\equiv&\left.\frac{\dln\sigma^2(R)}{\dln R}\right\vert_{\sigma=1}-3;~~~~ \mathcal{C}\equiv\left.\frac{\dsqln\sigma^2(R)}{\dln R^2}\right\vert_{\sigma=1},~~~~y\equiv\frac{k}{k_{\rm NL}},~~~~\sigma(k_{\rm NL}^{-1},z)=1\ea

As defined, $\sigma(R,z)$ is the root mean square dimensionless overdensity fluctuation within radius $R$ at redshift $z$, and it is used to quantify the nonlinear scale $k_{\rm NL}$, the scale on which $\sigma(R,z)$ reaches unity. The quantities $n_{\rm eff}$ and $\mathcal{C}$, which are the effective power index and running of the power spectrum respectively, have the shape information of the power spectrum.    

The functional form of $\alpha,~\beta,~\gamma,~\mu,~\nu,~a,~b,~c,~f_1,~f_2$ and $f_3$ are first given and calibrated by \citet{halofit}, and recently improved by a recalibration by \citet{halofitnew}. The recalibrated \hf~can fit $P(k)$ in $\Lambda$CDM very well (a 5\% precision at $k<1$ h/Mpc, $z\in[0,10]$) for a wide of range of cosmological parameters checked against the Coyote simulations \citep{Coyote1, Coyote2, Coyote3}. However, it fails to predict $P(k)$ in the HS model. In Figs \ref{fig:planck+wmap9} and \ref{fig:wmap7}, the \hf~and linear predictions of $\Delta_P$, which is the fractional difference in $P(k)$ as defined in Eq (\ref{eq:DP}), are shown in dash-dotted and dashed lines respectively. As shown, the linear theory overpredicts $\Delta_P$ in all cases, while {\tt Halofit} generally underpredicts $\Delta_P$ on quasi-nonlinear scales but overpredicts on fully nonlinear scales, especially for the F6 model, or all models at higher redshifts, where the screening effect is significant. For the F6 case, the prediction can even be negative, which is unphysical. This is partly because the {\tt Halofit} does not incorporate the screening mechanism at all. To generalise {\tt Halofit} for the HS model, the following requirements have to be met:

\begin{description}
\item[(A)]  It should well predict the power spectrum for a wide range of HS model parameter $f_{R0}$ and for various background cosmologies at various redshifts;  
\item[(B)]  When $|f_{R0}|\rightarrow0$, it should recover {\tt Halofit};
\item[(C)]  The screening effect must be included, \ie, for small field models ($|f_{R0}|\ll10^{-4}$), or at higher redshifts, the power should be suppressed compared to the {\tt Halofit} prediction on small scales; 
\item[(D)]  The suppression should decrease when $|f_{R0}|$ increases, or $z$ increases;
\item[(E)]  On large scale, the prediction should agree with the linear prediction;
\item[(F)]  On all scales, the prediction of $\Delta_P$ should not exceed the linear prediction; 
\item[(G)]  On all scales, $\Delta_P$ should be positive definite. 
\end{description}
   
Requirements (A) and (B) motivate the addition of new pieces of functions, which are proportional to $|f_{R0}|$, to the {\tt Halofit} formula. A multiplicative suppression term inversely proportional to $|f_{R0}|$ may satisfy (C, D), but it will break (B). Alternatively, we find that a multiplicative term inversely proportional to $\mathcal{D}$ works where \be\mathcal{D}\equiv \left\vert\frac{P(k)_{\rm HS}^{\rm lin.}}{P(k)_{\Lambda \rm {CDM}}^{\rm lin.}}-\frac{P(k)_{\rm HS}^{\hf}}{P(k)_{\Lambda {\rm CDM}}^{\hf}} \right\vert\ee and the superscript $^{\rm lin.}$ means the linear prediction. 

It is clear that, 
\begin{description}
\item[(I)]  $\mathcal{D}\rightarrow0$ when $|f_{R0}|\rightarrow0$, which guarantees a $\Lambda$CDM limit;
\item[(II)]  $\mathcal{D}\rightarrow0$ when $k\rightarrow0$ meaning that there is no suppression on large scales, which is what we want;
\item[(III)]  $\mathcal{D}$ can be large when $k$ is large meaning that the suppression increases with scale, which agrees with the trend of the chameleon screening; 
\item[(IV)]  $\mathcal{D}$ generally decreases when $|f_{R0}|$ increases in a wide range, which is what (D) requires. 
\item[(V)]  $\mathcal{D}$ can be easily calculated within {\tt MGCAMB} (\citealt{MGCAMB1,MGCAMB2}, see Appendix \ref{sec:apdxMGCAMB} for details of the implementation of the HS model in {\tt MGCAMB}) , making it a practically usable quantity for {\tt MGHalofit}. 
\end{description}

\begin{figure*}
  \begin{center}
  \includegraphics[scale=0.18]{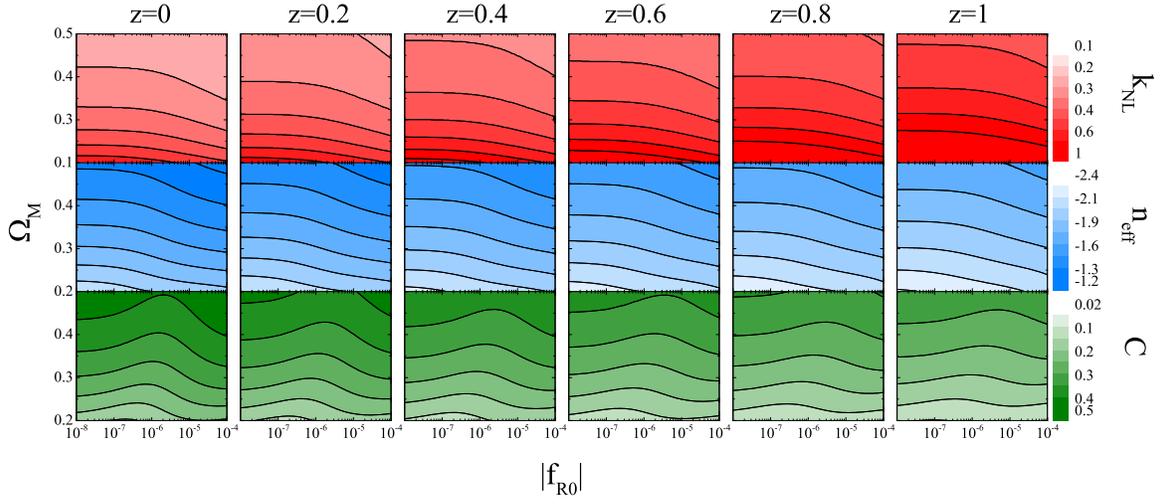}
  \end{center}
  \caption{The contour plots for $k_{\rm NL}$ (top panels), $n_{\rm eff}$ (middle) and $C$ (bottom) on the $(\Omega_{\rm M},~|f_{\rm R0}|)$ plane at various redshifts as illustrated in the figure.} \label{fig:contour}
\end{figure*}

Since the chameleon screening works in high density regions thus on small scales, we multiply this suppression factor on the nonlinear term $\Delta^2_{\rm H}$ to account for it, namely, \be\label{eq:damping} \Delta^2_{\rm H} \rightarrow \Delta^2_{\rm H}~{\rm exp}\left[\mathcal{D}\left(\vec{X}_{\mathcal{D}}\cdot\vec{Y}\right)\right] \ee where $\vec{X}_\mathcal{D}$ is a coefficient vector whose components are to be determined by $N$-body simulations, and \be \vec{Y}\equiv \left(1, n_{\rm eff}, n_{\rm eff}^2, \mathcal{C} \right)^T\ee Note that $\vec{Y}$ encodes the cosmology-dependence (see Eq (\ref{eq:shapepara}) for definitions of these quantities). In the HS model, the dependence of $k_{\rm NL}$, $n_{\rm eff}$ and $\mathcal{C}$ upon $\Omega_{\rm M}$ and $|f_{R0}|$ is shown in the contour plot Fig \ref{fig:contour}. As we can see, 

\begin{description}
\item[(I)] At a given redshift, say, $z=0$, for a given $\Omega_{\rm M}$, the nonlinear scale $y_{\rm NL}$ drops with $|f_{R0}|$ from some pivot point (\eg, at $\Omega_{\rm M}=0.3$, the transition is at $|f_{R0}|\sim10^{-6}$). This is because the linear power increases with $|f_{R0}|$, making the nonlinear scale larger for a larger $|f_{R0}|$; 
\item[(II)] For a given $|f_{R0}|$, $y_{\rm NL}$ drops with $\Omega_{\rm M}$ simply because more matter makes the clustering more significant; 
\item[(III)] Similarly, the effective power index $n_{\rm eff}$ increases with $|f_{R0}|$ or $\Omega_{\rm M}$ because more clustering makes the power spectrum less steep;   
\item[(IV)] The running of the power spectrum $\mathcal{C}$, which is essentially the gradient of $n_{\rm eff}$ follows a consistent trend, namely, when $n_{\rm eff}$ transits, a peak is produced in $\mathcal{C}$. 
\item[(V)] At higher redshifts, the overall dependence of $y_{\rm NL}, n_{\rm eff}$ and $\mathcal{C}$ stay largely unchanged, only with the difference in the amplitude.   
\end{description}

This is why we dot product the coefficients with $\vec{Y}$: we want the cosmology-dependence encoded in the damping term shown in Eq (\ref{eq:damping}). We have tested that adding this damping term can significantly improve the fit on small scales especially for the F6 model, but powers on quasi-nonlinear scales need to be boosted to further improve the fit. For this purpose,   
we generalise the functions in {\tt Halofit} by adding a term proportional to $|f_{R0}|$. Specifically,     
       
\remove{
In MGHalofit,

\be\tilde{\Delta}^2\equiv\frac{k^3\tilde{P}(k)}{2\pi^2}=\tilde{\Delta}^2_{\rm Q}+\tilde{\Delta}^2_{\rm H}\ee

\be\tilde{\Delta}^2_{\rm Q}(k)=\Delta^2_{\rm L}(k)\frac{[1+\tilde{\Delta}^2_{\rm L}(k)]^{\tilde{\beta}(n_{\rm eff},\mathcal{C},\mathcal{F})}}{1+\tilde{\alpha}(n_{\rm eff},\mathcal{C},\mathcal{F})\tilde{\Delta}^2_{\rm L}(k)}{\rm exp}[-f(y)]\ee 
\be\tilde{\Delta}^2_{\rm H}(k)=\frac{\tilde{\Delta}^{2'}_{\rm H}(k)\xi(n_{\rm eff},\mathcal{C},\mathcal{F})}{1+\tilde{\mu}(n_{\rm eff},\mathcal{C},\mathcal{F})/y+\tilde{\nu}(n_{\rm eff},\mathcal{C},\mathcal{F})/y^2}\ee

\be\tilde{\Delta}^{2'}_{\rm H}(k)=\frac{\tilde{a}(n_{\rm eff},\mathcal{C},\mathcal{F})y^{3f_1(\Omega)}}{1+\tilde{b}(n_{\rm eff},\mathcal{C},\mathcal{F})y^{f_2(\Omega)}
+[\tilde{c}(n_{\rm eff},\mathcal{C},\mathcal{F})f_3(\Omega)y]^{3-\tilde{\gamma}(n_{\rm eff},\mathcal{C},\mathcal{F})}} \ee

}

\ba\tilde{\Delta}^2_{\rm L}(k)&=&\Delta^2_{\rm L}(k)\left[1+ |f_{R0}|\left(\vec{X}_{\Delta}\cdot\vec{Y}\right)\right] \nonumber \\
\tilde{\alpha}&=&\alpha+
                                             |f_{R0}|\left(\vec{X}_{\alpha}\cdot\vec{Y}\right) \nonumber \\
 \tilde{\beta}&=&\beta+|f_{R0}|\left(\vec{X}_{\beta}\cdot\vec{Y}\right) \nonumber \\
 \tilde{\gamma}&=&\gamma+|f_{R0}|\left(\vec{X}_{\gamma}\cdot\vec{Y}\right)\nonumber \\
 {{\rm log}_{10}\tilde{a}}&=&{\rm log}_{10}\left[a+|f_{R0}|\left(\vec{X}_{a}\cdot\vec{Y}\right)\right]\nonumber \\
 {{\rm log}_{10}\tilde{b}}&=&{\rm log}_{10}\left[b+|f_{R0}|\left(\vec{X}_{b}\cdot\vec{Y}\right)\right]\nonumber \\
 {{\rm log}_{10}\tilde{c}}&=&{\rm log}_{10}\left[c+|f_{R0}|\left(\vec{X}_{c}\cdot\vec{Y}\right)\right]\nonumber \\
 {{\rm log}_{10}\tilde{\mu}}&=&{\rm log}_{10}\left[1+\mu+ |f_{R0}|\left(\vec{X}_{\mu}\cdot\vec{Y}\right)\right]\nonumber \\
 {{\rm log}_{10}\tilde{\nu}}&=&{\rm log}_{10}\left[\nu+ |f_{R0}|\left(\vec{X}_{\nu}\cdot\vec{Y}\right)\right] \ea

All this addition vanishes as $|f_{R0}|\rightarrow0$, yielding a $\Lambda$CDM limit. The functions $\alpha, \beta, \gamma, \mu, \nu, f_1, f_2$, $f_3$, $a, b$ and $c$ are given by \citet{halofitnew}, and the collection of coefficients $\vec{X}$ will be calibrated by the high-resolution $N$-body simulations described in the previous section. 


\subsection{Optimisation of the parameters}

To determine the new parameters $\vec{X}$, we minimise $\chi^2$, which is the quadratic difference between the model prediction and the simulation results of the fractional matter power spectrum. Specifically, 

\be \chi^2 = \sum_i [\Delta_P^{\rm sim.}(k_i)-\Delta_P^{\tt MGHalofit}(k_i)]^2 \ee where
\be \Delta_P^{\rm sim.}(k_i)\equiv \frac{P_{\rm HS}^{\rm sim.}(k_i)}{P_{\rm \Lambda CDM}^{\rm sim.}(k_i)}-1;~~~~\Delta_P^{\tt MGHalofit}(k_i)\equiv \frac{P_{\rm HS}^{\mgh}(k_i)}{P_{\rm \Lambda CDM}^{\tt Halofit}(k_i)}-1 \ee

where $\Delta_P^{\rm sim.}(k_i)$ and $\Delta_P^{\tt MGHalofit}(k_i)$ illustrate $\Delta_P$ in the $i$th bin (uniform in log $k$) calculated using the simulations and {\tt MGHalofit} respectively. The function minimisation was performed using the Powell's method \citep{Powell}, and the result is presented in Appendix \ref{sec:apdxMGHALOFIT}. 

The fitted result is shown in thick solid lines Figs \ref{fig:planck+wmap9} and \ref{fig:wmap7}. Compared to the linear prediction (dashed) and the \hf~prediction (dash-dotted), \mgh~agrees much better with the simulation result for all nine cosmologies (3 HS model $\times$ 3 background cosmologies) at various redshifts from $z=0$ to $z=1$. 

\subsection{Applicability of \mgh}

In this section, we shall quantify the accuracy of {\tt MGHalofit} and make a first application to estimate the detectability of the HS model using ongoing and upcoming redshift surveys.


\subsubsection{The accuracy of \mgh}

Note that the quantity $\Delta_P(k)$ shown in Figs \ref{fig:planck+wmap9} and \ref{fig:wmap7} is not a direct observable. For an imaging surveys such as the Dark Energy Survey (DES) and LSST \citep{LSST}, the observable is the shear angular correlation function or shear power spectrum, and the latter is essentially the matter power spectrum convolved with the lensing kernel. For redshift surveys, 
what is actually measured is the two-point correlation function in three-dimensions (3D), or the 3D galaxy power spectrum in redshift space, and the latter is the matter power spectrum in real space weighted by the galaxy bias and the redshift space distortion (RSD) correction. Therefore the matter power spectrum $P(k)$ is essentially the quantity directly related to observations. 

In our convention, the power spectrum for the HS model is related to that in $\Lambda$CDM via, \be P(k)_{\rm HS}^{\mgh}=\left[\Delta_P(k)+1\right]P(k)_{\rm \Lambda CDM}^{\hf}\ee where we have used {\mgh}~and \hf~to estimate the power spectrum in HS and $\Lambda$CDM models respectively. An error propagation gives (the $k$ dependence is dropped for brevity), \be \left[\frac{\sigma\left(P_{\rm HS}^{\mgh}\right)}{P_{\rm HS}^{\mgh}}\right]^2=\left[\frac{\sigma\left(P_{\rm \Lambda CDM}^{\hf}\right)}{P_{\rm \Lambda CDM}^{\hf}}\right]^2+\left[\frac{\sigma\left(\Delta_P\right)}{\Delta_P+1}\right]^2\ee          

The first term on the right hand side is the squared fractional accuracy of \hf, which is reported by \citet{halofitnew} to be below 5\% ($k\leqslant 1$ h/Mpc) and 10\% ($k\in(1, 10]$ h/Mpc) at $z\leqslant 3$ . The second term can be estimated by comparing the \mgh~prediction with the simulation result and in the worst case, $\sigma\left(\Delta_P\right)/(\Delta_P+1)=3\%$ ($k\leqslant 1$ h/Mpc) and 6\% ($k\in(1, 10]$ h/Mpc). This gives the accuracy of \mgh~as, \be\label{eq:accu} \frac{\sigma\left(P_{\rm HS}^{\mgh}\right)}{P_{\rm HS}^{\mgh}} \lesssim 6\%~(k\leqslant 1~{\rm h/Mpc});~~~~\lesssim 12\%~(k\in(1, 10]~{\rm h/Mpc}) \ee    

\subsubsection{A first application of \mgh}

In this section, we shall make a first application of \mgh~to estimate to what extent a HS model can be verified or falsified observationally. 


We first estimate the fractional difference in $P(k)$ using the Fisher matrix projection \citep{FKP,Seo}, \be\label{eq:FKP} \frac{\sigma_P(k)}{P(k)}=\frac{2\pi}{k\sqrt{V\Delta k}}\left(1+\frac{1}{\bar{n}P}\right) \ee where $V$ and $\bar{n}$ are the volume and the average galaxy number density of the surveys respectively. We make this forecast for an {\it Ongoing} and a {\it Future} redshift survey, whose survey parameters are listed in Table 2. The Ongoing survey is close to the Sloan Digital Sky Survey III's (SDSS-III) Baryon Oscillation Spectroscopic Survey (BOSS) survey (DR9) \citep{DR9}, and the Future survey is an idealised next-generation redshift survey similar to Dark Energy Spectroscopic. Instrument (DESI) \citep{DESI} and about a factor of 3-4 smaller than the Euclid spectroscopic survey \citep{Euclid}. 

\begin{table}[htdp]

\begin{center}
\begin{tabular}{c|cc}
\hline\hline
                                          & Ongoing survey & Future survey \\ 
                                          \hline
$z_{\rm eff}$                      &    0.6   &   1.0          \\
$V$ (Gpc$^3$ h$^{-3}$)                &    0.79   &    19.7            \\
$\bar{n}$ (h$^3$ Mpc$^{-3}$)       &      $3\times10^{-4}$      &     $4\times10^{-3}$             \\
\hline\hline
\end{tabular}
\end{center}
\label{tab:survey}
\caption{The survey parameters for an ongoing and a future survey.}
\end{table}%

\begin{figure*}
  \begin{center}
  \includegraphics[scale=0.3]{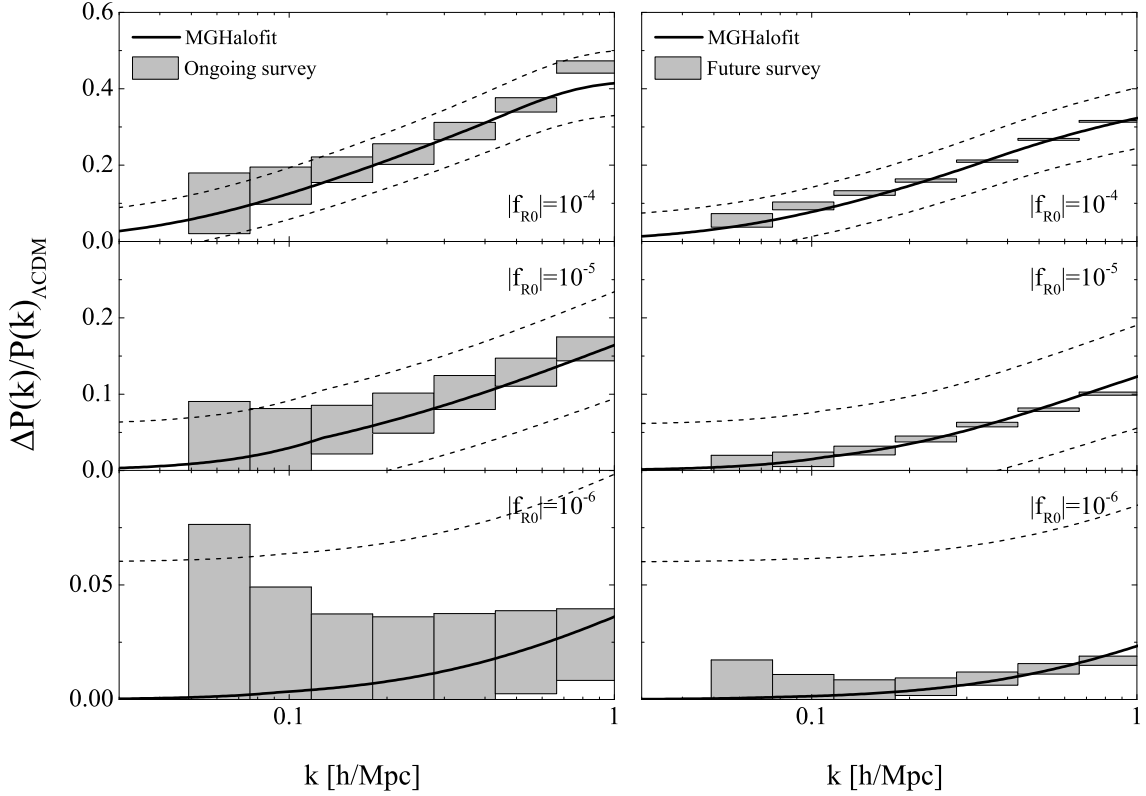}
  \end{center}
  \caption{The fractional difference in matter power spectrum for the Planck background cosmology. The power spectra are shown at $z=0.6$ and $z=1$ in the left and right panels respectively, which are the median redshifts for an ongoing and a future redshift survey respectively. See text for specifications of these surveys. The solid line shows the \mgh~prediction and the dashed line illustrate the 6\% error in $P(k)$. The error bars are based on a forecast using Eq (\ref{eq:FKP}) and the central values are taken from the $N$-body simulations.} \label{fig:cmass-euclid}
\end{figure*}

The result is shown in Fig \ref{fig:cmass-euclid}. The error bars are calculated using Eq (\ref{eq:FKP}) for ongoing and future surveys respectively and they are centered on $P(k)$ for HS fiducial models (F4, F5 and F6 from {top to bottom}) calculated using simulations. The solid curves show the \mgh~prediction with the dashed lines illustrate the 6\% error obtained in Eq (\ref{eq:accu}). Note that all curves and data points are rescaled using the \hf~prediction for the corresponding $\Lambda$CDM model just for the ease of visualisation. In all cases, \mgh~fits the simulation very well. 

Let us roughly estimate the detectability of the modification of gravity described by the HS model by calculating the $\chi^2$, \be \chi^2 = \sum_i \left[\frac{P(k_i)_{\rm HS}^{\rm sim.}-P(k_i)_{\rm \Lambda{CDM}}^{\rm sim.}}{\sigma_{\rm obs}(k_i)}\right]^2 \simeq \frac{\left[P(k_i)_{\rm HS}^{\mgh}-P(k_i)_{\rm \Lambda{CDM}}^{\hf}\right]^2}{\sigma_{\rm obs}^2(k_i)+\sigma_{\rm sys}^2(k_i)}\ee where in the second step we approximate the HS and $\Lambda$CDM simulations using \mgh~and \hf~respectively and this is why the systematic error $\sigma_{\rm sys}$ is added to the observational error $\sigma_{\rm obs}$ in quadrature. In this estimate we limit $k\leqslant 1$ h/Mpc and take $\sigma_{\rm sys}(k_i)=6\% \times P(k_i)_{\rm HS}^{\mgh}$. Under this setting, a future survey is able to detect the F4 and F5 models at the 6.9 and 2.3 $\sigma$ level respectively while the F6 model will never be detected. This is easy to understand: the signal of the F6 model is even below the 5\% accuracy of \hf. Note that this is a rough estimate where we ignore the uncertainties of the galaxy bias and RSD (for RSD in the HS model, see \citealt{HSRSD}), which are below the level of $\sigma_{\rm obs}$ though. We also ignore the degeneracy with other cosmological parameters. We will perform a detailed cosmological forecast using \mgh~in a future publication. 

\begin{figure*}
  \begin{center}
  \includegraphics[scale=0.35]{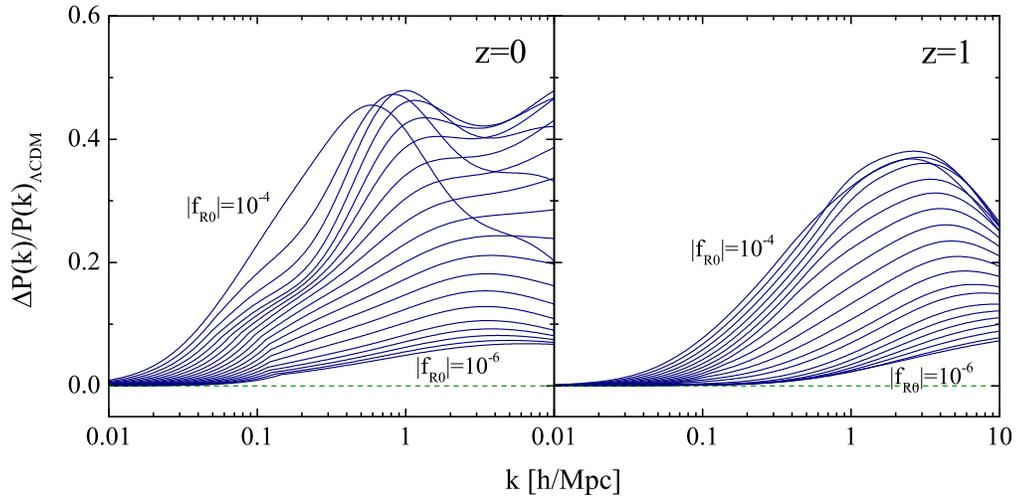}
  \end{center}
  \caption{The fractional difference in matter power spectrum $\Delta_P(k)$ calculated using \mgh. The curves from {top to bottom} are for 20 different $|f_{R0}|$ values ranging from $10^{-4}$ to $10^{-6}$ (uniform logarithmically). The left and right panels show the result at $z=0$ and $z=1$ respectively, and the horizontal green dashed line illustrates $\Delta_P(k)=0$ for a reference.} \label{fig:dPoP_20fR0}
\end{figure*}

Note that \mgh~works for an arbitrary $|f_{R0}|\in[10^{-6},10^{-4}]$ below redshift $z=1$, and Fig \ref{fig:dPoP_20fR0} shows the result for $20$ HS models with $|f_{R0}|$ logarithmically uniform from $10^{-6}$ to $10^{-4}$ at redshifts $z=0$ and $z=1$. Given the current constraint on $|f_{R0}|$, which is $|f_{R0}|\lesssim 10^{-4}$ \citep{HSobs8} and the redshift range of future surveys, \mgh~is sufficient for observational tests of the HS model. 

\section{Conclusion and Discussion}

In this work, we develop a new fitting formula \mgh~to calculate the nonlinear matter power spectrum for the Hu-Sawicki $f(R)$ model. The fitting formula is developed by generalising the \hf~fitting formula to include the chameleon screening mechanism, and it was calibrated using a suite of high-resolution HS $N$-body simulations with various model and cosmological parameters. Compared to the \hf~prescription, \mgh~significantly improve the fit, namely, \mgh~reaches an accuracy of 6\% and 12\% at $k\leqslant 1$ h/Mpc and $k\in(1,10]$ h/Mpc respectively below redshift 1. 

\mgh~can be used for parameter constraints for the HS model using a large class of future imaging, spectroscopic and 21$cm$ surveys including the Dark Energy Survey (DES) \footnote{\url{https://www.darkenergysurvey.org}}, the Large Synoptic Survey Telescope (LSST) \footnote{\url{http://www.lsst.org/lsst/}} \citep{LSST}, DESI \citep{DESI}, Euclid \footnote{\url{http://sci.esa.int/euclid/}} \citep{Euclid}, Square Kilometer Array (SKA) \footnote{\url{https://www.skatelescope.org/}} and so on. 

We make a first application of \mgh~to estimate to what extent the HS model can be verified or falsified using the ongoing and forthcoming redshift surveys, and find that a future redshift survey is able to detect the F4 and F5 models at the 6.9 and 2.3 $\sigma$ levels respectively. It is difficult to detect a model with $|f_{R0}|<10^{-5}$ using $P(k)$ even up to nonlinear scales because we are limited not only by the accuracy of the fitting formulae, but also by the complicated astrophysical systematics on such scales. However, the constraint can be further improved using alternative approaches, \eg~searching for the environmental dependence of the screening \citep{HSenv}, performing astrophysical tests of MG using galaxy dynamics \citep{HSastro1,HSastro2,HSastro3,HSastro4,HSspin}, as well as measuring the cluster density profiles \citep{HSobs5}.



\section*{Acknowledgements}

I thank Pedro Ferreira, Bhuvnesh Jain, Kazuya Koyama, Baojiu Li and Lucas Lombriser for discussions, and the Euclid Science Coordinators for correspondence regarding the specification of future spectroscopic surveys. This work is supported by the $1000$ Young Talents program in China, by the 973 Program grant No. 2013CB837900, NSFC grant No. 11261140641, and CAS grant No. KJZD-EW-T01, and by University of Portsmouth. All numeric calculations were performed on the SCIAMA supercomputer at University of Portsmouth. 

\bibliographystyle{apj}
\bibliography{resubmit}

\begin{thebibliography}{}
\expandafter\ifx\csname natexlab\endcsname\relax\def\natexlab#1{#1}\fi

\bibitem[{{Anderson} {et~al.}(2012){Anderson}, {Aubourg}, {Bailey}, {Bizyaev},
  {Blanton}, {Bolton}, {Brinkmann}, {Brownstein}, {Burden}, {Cuesta}, {da
  Costa}, {Dawson}, {de Putter}, {Eisenstein}, {Gunn}, {Guo}, {Hamilton},
  {Harding}, {Ho}, {Honscheid}, {Kazin}, {Kirkby}, {Kneib}, {Labatie},
  {Loomis}, {Lupton}, {Malanushenko}, {Malanushenko}, {Mandelbaum}, {Manera},
  {Maraston}, {McBride}, {Mehta}, {Mena}, {Montesano}, {Muna}, {Nichol},
  {Nuza}, {Olmstead}, {Oravetz}, {Padmanabhan}, {Palanque-Delabrouille}, {Pan},
  {Parejko}, {P{\^a}ris}, {Percival}, {Petitjean}, {Prada}, {Reid}, {Roe},
  {Ross}, {Ross}, {Samushia}, {S{\'a}nchez}, {Schlegel}, {Schneider},
  {Sc{\'o}ccola}, {Seo}, {Sheldon}, {Simmons}, {Skibba}, {Strauss}, {Swanson},
  {Thomas}, {Tinker}, {Tojeiro}, {Maga{\~n}a}, {Verde}, {Wagner}, {Wake},
  {Weaver}, {Weinberg}, {White}, {Xu}, {Y{\`e}che}, {Zehavi}, \& {Zhao}}]{DR9}
{Anderson}, L., {Aubourg}, E., {Bailey}, S., {et~al.} 2012, \mnras, 427, 3435

\bibitem[{{Baker} {et~al.}(2013{\natexlab{a}}){Baker}, {Ferreira}, \&
  {Skordis}}]{MGpara8}
{Baker}, T., {Ferreira}, P.~G., \& {Skordis}, C. 2013{\natexlab{a}}, ArXiv
  e-prints, arXiv:1310.1086

\bibitem[{{Baker} {et~al.}(2013{\natexlab{b}}){Baker}, {Ferreira}, \&
  {Skordis}}]{MGpara3}
---. 2013{\natexlab{b}}, \prd, 87, 024015

\bibitem[{{Baker} {et~al.}(2011){Baker}, {Ferreira}, {Skordis}, \&
  {Zuntz}}]{MGpara1}
{Baker}, T., {Ferreira}, P.~G., {Skordis}, C., \& {Zuntz}, J. 2011, \prd, 84,
  124018

\bibitem[{{Bean} \& {Tangmatitham}(2010)}]{MGpara5}
{Bean}, R., \& {Tangmatitham}, M. 2010, \prd, 81, 083534

\bibitem[{{Bertschinger}(1995)}]{COSMICS}
{Bertschinger}, E. 1995, ArXiv Astrophysics e-prints, astro-ph/9506070

\bibitem[{{Bertschinger} \& {Zukin}(2008)}]{BZ}
{Bertschinger}, E., \& {Zukin}, P. 2008, \prd, 78, 024015

\bibitem[{{Bird} {et~al.}(2012){Bird}, {Viel}, \& {Haehnelt}}]{halofitmnu}
{Bird}, S., {Viel}, M., \& {Haehnelt}, M.~G. 2012, \mnras, 420, 2551

\bibitem[{{Brax} {et~al.}(2012){Brax}, {Davis}, {Li}, {Winther}, \&
  {Zhao}}]{sim:symmetron+dilaton}
{Brax}, P., {Davis}, A.-C., {Li}, B., {Winther}, H.~A., \& {Zhao}, G.-B. 2012,
  \jcap, 10, 2

\bibitem[{{Brax} {et~al.}(2013){Brax}, {Davis}, {Li}, {Winther}, \&
  {Zhao}}]{sim:chameleon}
---. 2013, \jcap, 4, 29

\bibitem[{{Cabr{\'e}} {et~al.}(2012){Cabr{\'e}}, {Vikram}, {Zhao}, {Jain}, \&
  {Koyama}}]{HSastro2}
{Cabr{\'e}}, A., {Vikram}, V., {Zhao}, G.-B., {Jain}, B., \& {Koyama}, K. 2012,
  \jcap, 7, 34

\bibitem[{{Clifton} {et~al.}(2012){Clifton}, {Ferreira}, {Padilla}, \&
  {Skordis}}]{MGreview}
{Clifton}, T., {Ferreira}, P.~G., {Padilla}, A., \& {Skordis}, C. 2012,
  \physrep, 513, 1

\bibitem[{{Colombi} {et~al.}(2009){Colombi}, {Jaffe}, {Novikov}, \&
  {Pichon}}]{POWMES}
{Colombi}, S., {Jaffe}, A., {Novikov}, D., \& {Pichon}, C. 2009, \mnras, 393,
  511

\bibitem[{{Cooray} \& {Sheth}(2002)}]{haloreview}
{Cooray}, A., \& {Sheth}, R. 2002, \physrep, 372, 1

\bibitem[{{Feldman} {et~al.}(1994){Feldman}, {Kaiser}, \& {Peacock}}]{FKP}
{Feldman}, H.~A., {Kaiser}, N., \& {Peacock}, J.~A. 1994, \apj, 426, 23

\bibitem[{{Giannantonio} {et~al.}(2010){Giannantonio}, {Martinelli},
  {Silvestri}, \& {Melchiorri}}]{HSobs4}
{Giannantonio}, T., {Martinelli}, M., {Silvestri}, A., \& {Melchiorri}, A.
  2010, \jcap, 4, 30

\bibitem[{{Hamilton} {et~al.}(1991){Hamilton}, {Kumar}, {Lu}, \&
  {Matthews}}]{HKLM}
{Hamilton}, A.~J.~S., {Kumar}, P., {Lu}, E., \& {Matthews}, A. 1991, \apjl,
  374, L1

\bibitem[{{Heitmann} {et~al.}(2009){Heitmann}, {Higdon}, {White}, {Habib},
  {Williams}, {Lawrence}, \& {Wagner}}]{Coyote2}
{Heitmann}, K., {Higdon}, D., {White}, M., {et~al.} 2009, \apj, 705, 156

\bibitem[{{Heitmann} {et~al.}(2010){Heitmann}, {White}, {Wagner}, {Habib}, \&
  {Higdon}}]{Coyote1}
{Heitmann}, K., {White}, M., {Wagner}, C., {Habib}, S., \& {Higdon}, D. 2010,
  \apj, 715, 104

\bibitem[{{Hinshaw} {et~al.}(2013){Hinshaw}, {Larson}, {Komatsu}, {Spergel},
  {Bennett}, {Dunkley}, {Nolta}, {Halpern}, {Hill}, {Odegard}, {Page}, {Smith},
  {Weiland}, {Gold}, {Jarosik}, {Kogut}, {Limon}, {Meyer}, {Tucker}, {Wollack},
  \& {Wright}}]{WMAP9}
{Hinshaw}, G., {Larson}, D., {Komatsu}, E., {et~al.} 2013, \apjs, 208, 19

\bibitem[{{Hojjati} {et~al.}(2011){Hojjati}, {Pogosian}, \& {Zhao}}]{MGCAMB2}
{Hojjati}, A., {Pogosian}, L., \& {Zhao}, G.-B. 2011, \jcap, 8, 5

\bibitem[{{Hu} \& {Sawicki}(2007{\natexlab{a}})}]{HS}
{Hu}, W., \& {Sawicki}, I. 2007{\natexlab{a}}, \prd, 76, 064004

\bibitem[{{Hu} \& {Sawicki}(2007{\natexlab{b}})}]{ppf}
---. 2007{\natexlab{b}}, \prd, 76, 104043

\bibitem[{{Jain} \& {Bertschinger}(1998)}]{PD96fail3}
{Jain}, B., \& {Bertschinger}, E. 1998, \apj, 509, 517

\bibitem[{{Jain} \& {Khoury}(2010)}]{HSobsrev}
{Jain}, B., \& {Khoury}, J. 2010, Annals of Physics, 325, 1479

\bibitem[{{Jain} {et~al.}(1995){Jain}, {Mo}, \& {White}}]{JMW95}
{Jain}, B., {Mo}, H.~J., \& {White}, S.~D.~M. 1995, \mnras, 276, L25

\bibitem[{{Jain} \& {VanderPlas}(2011)}]{HSastro1}
{Jain}, B., \& {VanderPlas}, J. 2011, \jcap, 10, 32

\bibitem[{{Jain} {et~al.}(2013){Jain}, {Vikram}, \& {Sakstein}}]{HSastro4}
{Jain}, B., {Vikram}, V., \& {Sakstein}, J. 2013, \apj, 779, 39

\bibitem[{{Jennings} {et~al.}(2012){Jennings}, {Baugh}, {Li}, {Zhao}, \&
  {Koyama}}]{HSRSD}
{Jennings}, E., {Baugh}, C.~M., {Li}, B., {Zhao}, G.-B., \& {Koyama}, K. 2012,
  \mnras, 425, 2128

\bibitem[{{Khoury} \& {Weltman}(2004)}]{chameleon}
{Khoury}, J., \& {Weltman}, A. 2004, \prd, 69, 044026

\bibitem[{{Knebe} {et~al.}(2001){Knebe}, {Green}, \& {Binney}}]{MLAPM}
{Knebe}, A., {Green}, A., \& {Binney}, J. 2001, \mnras, 325, 845

\bibitem[{{Komatsu} {et~al.}(2011){Komatsu}, {Smith}, {Dunkley}, {Bennett},
  {Gold}, {Hinshaw}, {Jarosik}, {Larson}, {Nolta}, {Page}, {Spergel},
  {Halpern}, {Hill}, {Kogut}, {Limon}, {Meyer}, {Odegard}, {Tucker}, {Weiland},
  {Wollack}, \& {Wright}}]{WMAP7}
{Komatsu}, E., {Smith}, K.~M., {Dunkley}, J., {et~al.} 2011, \apjs, 192, 18

\bibitem[{{Koyama} {et~al.}(2009){Koyama}, {Taruya}, \& {Hiramatsu}}]{KK09}
{Koyama}, K., {Taruya}, A., \& {Hiramatsu}, T. 2009, \prd, 79, 123512

\bibitem[{{Lam} {et~al.}(2012){Lam}, {Nishimichi}, {Schmidt}, \&
  {Takada}}]{HSobs6}
{Lam}, T.~Y., {Nishimichi}, T., {Schmidt}, F., \& {Takada}, M. 2012, Physical
  Review Letters, 109, 051301

\bibitem[{{Laureijs} {et~al.}(2011){Laureijs}, {Amiaux}, {Arduini},
  {Augu{\`e}res}, {Brinchmann}, {Cole}, {Cropper}, {Dabin}, {Duvet}, {Ealet},
  \& et~al.}]{Euclid}
{Laureijs}, R., {Amiaux}, J., {Arduini}, S., {et~al.} 2011, ArXiv e-prints,
  arXiv:1110.3193

\bibitem[{{Lawrence} {et~al.}(2010){Lawrence}, {Heitmann}, {White}, {Higdon},
  {Wagner}, {Habib}, \& {Williams}}]{Coyote3}
{Lawrence}, E., {Heitmann}, K., {White}, M., {et~al.} 2010, \apj, 713, 1322

\bibitem[{{Lee} {et~al.}(2013){Lee}, {Zhao}, {Li}, \& {Koyama}}]{HSspin}
{Lee}, J., {Zhao}, G.-B., {Li}, B., \& {Koyama}, K. 2013, \apj, 763, 28

\bibitem[{{Levi} {et~al.}(2013){Levi}, {Bebek}, {Beers}, {Blum}, {Cahn},
  {Eisenstein}, {Flaugher}, {Honscheid}, {Kron}, {Lahav}, {McDonald}, {Roe},
  {Schlegel}, \& {representing the DESI collaboration}}]{DESI}
{Levi}, M., {Bebek}, C., {Beers}, T., {et~al.} 2013, ArXiv e-prints,
  arXiv:1308.0847

\bibitem[{{Li} {et~al.}(2013){Li}, {Hellwing}, {Koyama}, {Zhao}, {Jennings}, \&
  {Baugh}}]{HSPk}
{Li}, B., {Hellwing}, W.~A., {Koyama}, K., {et~al.} 2013, \mnras, 428, 743

\bibitem[{{Li} {et~al.}(2012){Li}, {Zhao}, {Teyssier}, \& {Koyama}}]{ECOSMOG}
{Li}, B., {Zhao}, G.-B., {Teyssier}, R., \& {Koyama}, K. 2012, \jcap, 1, 51

\bibitem[{{Linder}(2005)}]{MGpara7}
{Linder}, E.~V. 2005, \prd, 72, 043529

\bibitem[{{Lombriser} {et~al.}(2013){Lombriser}, {Koyama}, \& {Li}}]{Pkhalo}
{Lombriser}, L., {Koyama}, K., \& {Li}, B. 2013, ArXiv e-prints,
  arXiv:1312.1292

\bibitem[{{Lombriser} {et~al.}(2012{\natexlab{a}}){Lombriser}, {Schmidt},
  {Baldauf}, {Mandelbaum}, {Seljak}, \& {Smith}}]{HSobs5}
{Lombriser}, L., {Schmidt}, F., {Baldauf}, T., {et~al.} 2012{\natexlab{a}},
  \prd, 85, 102001

\bibitem[{{Lombriser} {et~al.}(2012{\natexlab{b}}){Lombriser}, {Slosar},
  {Seljak}, \& {Hu}}]{HSobs3}
{Lombriser}, L., {Slosar}, A., {Seljak}, U., \& {Hu}, W. 2012{\natexlab{b}},
  \prd, 85, 124038

\bibitem[{{LSST Science Collaboration} {et~al.}(2009){LSST Science
  Collaboration}, {Abell}, {Allison}, {Anderson}, {Andrew}, {Angel}, {Armus},
  {Arnett}, {Asztalos}, {Axelrod}, \& et~al.}]{LSST}
{LSST Science Collaboration}, {Abell}, P.~A., {Allison}, J., {et~al.} 2009,
  ArXiv e-prints, arXiv:0912.0201

\bibitem[{{Mak} {et~al.}(2012){Mak}, {Pierpaoli}, {Schmidt}, \&
  {Macellari}}]{HSobs7}
{Mak}, D.~S.~Y., {Pierpaoli}, E., {Schmidt}, F., \& {Macellari}, N. 2012, \prd,
  85, 123513

\bibitem[{{McClelland} \& {Silk}(1977)}]{halo7}
{McClelland}, J., \& {Silk}, J. 1977, \apj, 216, 665

\bibitem[{{Mo} {et~al.}(1997{\natexlab{a}}){Mo}, {Jing}, \&
  {Borner}}]{PD96fail1}
{Mo}, H.~J., {Jing}, Y.~P., \& {Borner}, G. 1997{\natexlab{a}}, \mnras, 286,
  979

\bibitem[{{Mo} {et~al.}(1997{\natexlab{b}}){Mo}, {Jing}, \& {White}}]{halo2}
{Mo}, H.~J., {Jing}, Y.~P., \& {White}, S.~D.~M. 1997{\natexlab{b}}, \mnras,
  284, 189

\bibitem[{{Mo} \& {White}(1996)}]{halo1}
{Mo}, H.~J., \& {White}, S.~D.~M. 1996, \mnras, 282, 347

\bibitem[{{Noller} {et~al.}(2013){Noller}, {von Braun-Bates}, \&
  {Ferreira}}]{quasistatic}
{Noller}, J., {von Braun-Bates}, F., \& {Ferreira}, P.~G. 2013, ArXiv e-prints,
  arXiv:1310.3266

\bibitem[{{Oyaizu}(2008)}]{Oyaizu1}
{Oyaizu}, H. 2008, \prd, 78, 123523

\bibitem[{{Oyaizu} {et~al.}(2008){Oyaizu}, {Lima}, \& {Hu}}]{Oyaizu2}
{Oyaizu}, H., {Lima}, M., \& {Hu}, W. 2008, \prd, 78, 123524

\bibitem[{{Peacock} \& {Dodds}(1994)}]{PD94}
{Peacock}, J.~A., \& {Dodds}, S.~J. 1994, \mnras, 267, 1020

\bibitem[{{Peacock} \& {Dodds}(1996)}]{PD96}
---. 1996, \mnras, 280, L19

\bibitem[{{Peebles}(1974)}]{halo6}
{Peebles}, P.~J.~E. 1974, \aap, 32, 197

\bibitem[{{Perlmutter} {et~al.}(1999){Perlmutter}, {Aldering}, {Goldhaber},
  {Knop}, {Nugent}, {Castro}, {Deustua}, {Fabbro}, {Goobar}, {Groom}, {Hook},
  {Kim}, {Kim}, {Lee}, {Nunes}, {Pain}, {Pennypacker}, {Quimby}, {Lidman},
  {Ellis}, {Irwin}, {McMahon}, {Ruiz-Lapuente}, {Walton}, {Schaefer}, {Boyle},
  {Filippenko}, {Matheson}, {Fruchter}, {Panagia}, {Newberg}, {Couch}, \&
  {Supernova Cosmology Project}}]{Perlmutter}
{Perlmutter}, S., {Aldering}, G., {Goldhaber}, G., {et~al.} 1999, \apj, 517,
  565

\bibitem[{{Planck Collaboration} {et~al.}(2013){Planck Collaboration}, {Ade},
  {Aghanim}, {Armitage-Caplan}, {Arnaud}, {Ashdown}, {Atrio-Barandela},
  {Aumont}, {Baccigalupi}, {Banday}, \& et~al.}]{Planck}
{Planck Collaboration}, {Ade}, P.~A.~R., {Aghanim}, N., {et~al.} 2013, ArXiv
  e-prints, arXiv:1303.5076

\bibitem[{{Pogosian} \& {Silvestri}(2008)}]{PS}
{Pogosian}, L., \& {Silvestri}, A. 2008, \prd, 77, 023503

\bibitem[{{Pogosian} {et~al.}(2010){Pogosian}, {Silvestri}, {Koyama}, \&
  {Zhao}}]{pogosian}
{Pogosian}, L., {Silvestri}, A., {Koyama}, K., \& {Zhao}, G.-B. 2010, \prd, 81,
  104023

\bibitem[{{Powell}(1964)}]{Powell}
{Powell}, M. J.~D. 1964, The Computer Journal, 7, 155

\bibitem[{{Prunet} {et~al.}(2008){Prunet}, {Pichon}, {Aubert}, {Pogosyan},
  {Teyssier}, \& {Gottloeber}}]{MPgrafic}
{Prunet}, S., {Pichon}, C., {Aubert}, D., {et~al.} 2008, \apjs, 178, 179

\bibitem[{{Puchwein} {et~al.}(2013){Puchwein}, {Baldi}, \&
  {Springel}}]{MGGadget}
{Puchwein}, E., {Baldi}, M., \& {Springel}, V. 2013, \mnras, 436, 348

\bibitem[{{Reyes} {et~al.}(2010){Reyes}, {Mandelbaum}, {Seljak}, {Baldauf},
  {Gunn}, {Lombriser}, \& {Smith}}]{HSobs10}
{Reyes}, R., {Mandelbaum}, R., {Seljak}, U., {et~al.} 2010, \nat, 464, 256

\bibitem[{{Riess} {et~al.}(1998){Riess}, {Filippenko}, {Challis},
  {Clocchiatti}, {Diercks}, {Garnavich}, {Gilliland}, {Hogan}, {Jha},
  {Kirshner}, {Leibundgut}, {Phillips}, {Reiss}, {Schmidt}, {Schommer},
  {Smith}, {Spyromilio}, {Stubbs}, {Suntzeff}, \& {Tonry}}]{Riess}
{Riess}, A.~G., {Filippenko}, A.~V., {Challis}, P., {et~al.} 1998, \aj, 116,
  1009

\bibitem[{{Schmidt} {et~al.}(2009{\natexlab{a}}){Schmidt}, {Lima}, {Oyaizu}, \&
  {Hu}}]{Oyaizu3}
{Schmidt}, F., {Lima}, M., {Oyaizu}, H., \& {Hu}, W. 2009{\natexlab{a}}, \prd,
  79, 083518

\bibitem[{{Schmidt} {et~al.}(2009{\natexlab{b}}){Schmidt}, {Vikhlinin}, \&
  {Hu}}]{HSobs8}
{Schmidt}, F., {Vikhlinin}, A., \& {Hu}, W. 2009{\natexlab{b}}, \prd, 80,
  083505

\bibitem[{{Seo} \& {Eisenstein}(2007)}]{Seo}
{Seo}, H.-J., \& {Eisenstein}, D.~J. 2007, \apj, 665, 14

\bibitem[{{Sheth} \& {Jain}(1997)}]{halo8}
{Sheth}, R.~K., \& {Jain}, B. 1997, \mnras, 285, 231

\bibitem[{{Sheth} \& {Lemson}(1999)}]{halo3}
{Sheth}, R.~K., \& {Lemson}, G. 1999, \mnras, 304, 767

\bibitem[{{Sheth} {et~al.}(2001){Sheth}, {Mo}, \& {Tormen}}]{halo5}
{Sheth}, R.~K., {Mo}, H.~J., \& {Tormen}, G. 2001, \mnras, 323, 1

\bibitem[{{Sheth} \& {Tormen}(1999)}]{halo4}
{Sheth}, R.~K., \& {Tormen}, G. 1999, \mnras, 308, 119

\bibitem[{{Smith} {et~al.}(1998){Smith}, {Klypin}, {Gross}, {Primack}, \&
  {Holtzman}}]{PD96fail2}
{Smith}, C.~C., {Klypin}, A., {Gross}, M.~A.~K., {Primack}, J.~R., \&
  {Holtzman}, J. 1998, \mnras, 297, 910

\bibitem[{{Smith} {et~al.}(2003){Smith}, {Peacock}, {Jenkins}, {White},
  {Frenk}, {Pearce}, {Thomas}, {Efstathiou}, \& {Couchman}}]{halofit}
{Smith}, R.~E., {Peacock}, J.~A., {Jenkins}, A., {et~al.} 2003, \mnras, 341,
  1311

\bibitem[{{Song} {et~al.}(2007{\natexlab{a}}){Song}, {Hu}, \&
  {Sawicki}}]{HSobs1}
{Song}, Y.-S., {Hu}, W., \& {Sawicki}, I. 2007{\natexlab{a}}, \prd, 75, 044004

\bibitem[{{Song} {et~al.}(2007{\natexlab{b}}){Song}, {Peiris}, \&
  {Hu}}]{HSobs2}
{Song}, Y.-S., {Peiris}, H., \& {Hu}, W. 2007{\natexlab{b}}, \prd, 76, 063517

\bibitem[{{Song} {et~al.}(2011){Song}, {Zhao}, {Bacon}, {Koyama}, {Nichol}, \&
  {Pogosian}}]{MGpara4}
{Song}, Y.-S., {Zhao}, G.-B., {Bacon}, D., {et~al.} 2011, \prd, 84, 083523

\bibitem[{{Springel}(2005)}]{GADGET}
{Springel}, V. 2005, \mnras, 364, 1105

\bibitem[{{Takahashi} {et~al.}(2012){Takahashi}, {Sato}, {Nishimichi},
  {Taruya}, \& {Oguri}}]{halofitnew}
{Takahashi}, R., {Sato}, M., {Nishimichi}, T., {Taruya}, A., \& {Oguri}, M.
  2012, \apj, 761, 152

\bibitem[{{Teyssier}(2002)}]{Ramses}
{Teyssier}, R. 2002, \aap, 385, 337

\bibitem[{{Thomas} {et~al.}(2011){Thomas}, {Appleby}, \& {Weller}}]{MGpara6}
{Thomas}, S.~A., {Appleby}, S.~A., \& {Weller}, J. 2011, \jcap, 3, 36

\bibitem[{{Van Waerbeke} {et~al.}(2001){Van Waerbeke}, {Mellier}, {Radovich},
  {Bertin}, {Dantel-Fort}, {McCracken}, {Le F{\`e}vre}, {Foucaud},
  {Cuillandre}, {Erben}, {Jain}, {Schneider}, {Bernardeau}, \&
  {Fort}}]{PD96fail4}
{Van Waerbeke}, L., {Mellier}, Y., {Radovich}, M., {et~al.} 2001, \aap, 374,
  757

\bibitem[{{Vikram} {et~al.}(2013){Vikram}, {Cabr{\'e}}, {Jain}, \&
  {VanderPlas}}]{HSastro3}
{Vikram}, V., {Cabr{\'e}}, A., {Jain}, B., \& {VanderPlas}, J.~T. 2013, \jcap,
  8, 20

\bibitem[{{Weinberg} {et~al.}(2013){Weinberg}, {Mortonson}, {Eisenstein},
  {Hirata}, {Riess}, \& {Rozo}}]{DEreview}
{Weinberg}, D.~H., {Mortonson}, M.~J., {Eisenstein}, D.~J., {et~al.} 2013,
  \physrep, 530, 87

\bibitem[{{Zhang} {et~al.}(2007){Zhang}, {Liguori}, {Bean}, \&
  {Dodelson}}]{HSobs9}
{Zhang}, P., {Liguori}, M., {Bean}, R., \& {Dodelson}, S. 2007, Physical Review
  Letters, 99, 141302

\bibitem[{{Zhao} {et~al.}(2011{\natexlab{a}}){Zhao}, {Li}, \&
  {Koyama}}]{Zhao11}
{Zhao}, G.-B., {Li}, B., \& {Koyama}, K. 2011{\natexlab{a}}, \prd, 83, 044007

\bibitem[{{Zhao} {et~al.}(2011{\natexlab{b}}){Zhao}, {Li}, \& {Koyama}}]{HSenv}
---. 2011{\natexlab{b}}, Physical Review Letters, 107, 071303

\bibitem[{{Zhao} {et~al.}(2009){Zhao}, {Pogosian}, {Silvestri}, \&
  {Zylberberg}}]{MGCAMB1}
{Zhao}, G.-B., {Pogosian}, L., {Silvestri}, A., \& {Zylberberg}, J. 2009, \prd,
  79, 083513

\bibitem[{{Zuntz} {et~al.}(2012){Zuntz}, {Baker}, {Ferreira}, \&
  {Skordis}}]{MGpara2}
{Zuntz}, J., {Baker}, T., {Ferreira}, P.~G., \& {Skordis}, C. 2012, \jcap, 6,
  32

\end{thebibliography}

\newpage

\appendix

\section{MGHALOFIT Fitting formula}
\label{sec:apdxMGHALOFIT}
We extend \hf~by adding new terms for the HS model and calibrate these terms using simulations. In what follows, we shall present the full \mgh~fitting formula and the numeric value of the coefficients therein. Note that the quantities with tildes denote the revised quantity while those without a tilde represent the quantity in the newly calibrated \hf~presented in the Appendix of \citet{halofitnew}. 

\be\tilde{\Delta}^2\equiv\frac{k^3P(k)_{\rm HS}^{\mgh}}{2\pi^2}=\tilde{\Delta}^2_{\rm Q}+\tilde{\Delta}^2_{\rm H}\ee

\be\tilde{\Delta}^2_{\rm Q}(k)=\Delta^2_{\rm L}(k)\frac{[1+\tilde{\Delta}^2_{\rm L}(k)]^{\tilde{\beta}(n_{\rm eff},\mathcal{C},\mathcal{F})}}{1+\tilde{\alpha}(n_{\rm eff},\mathcal{C},\mathcal{F})\tilde{\Delta}^2_{\rm L}(k)}{\rm exp}[-f(y)]\ee 
\be\tilde{\Delta}^2_{\rm H}(k)=\frac{\tilde{\Delta}^{2'}_{\rm H}(k)\xi(n_{\rm eff},\mathcal{C},\mathcal{F})}{1+\tilde{\mu}(n_{\rm eff},\mathcal{C},\mathcal{F})/y+\tilde{\nu}(n_{\rm eff},\mathcal{C},\mathcal{F})/y^2}\ee

\be\tilde{\Delta}^{2'}_{\rm H}(k)=\frac{\tilde{a}(n_{\rm eff},\mathcal{C},\mathcal{F})y^{3f_1(\Omega)}}{1+\tilde{b}(n_{\rm eff},\mathcal{C},\mathcal{F})y^{f_2(\Omega)}
+[\tilde{c}(n_{\rm eff},\mathcal{C},\mathcal{F})f_3(\Omega)y]^{3-\tilde{\gamma}(n_{\rm eff},\mathcal{C},\mathcal{F})}} \ee


\ba\tilde{\Delta}^2_{\rm L}(k)&=&\Delta^2_{\rm L}(k)\left[1+ \mathcal{F}\left(x_1+x_2n_{\rm eff}+x_3\mathcal{C}\right)\right] \nonumber \\
\tilde{\alpha}&=&\alpha+
                                             \mathcal{F}\left(x_4+x_5n_{\rm eff}+x_6n_{\rm eff}^2+x_7\mathcal{C} \right) \nonumber \\
 \tilde{\beta}&=&\beta+\mathcal{F}\left(x_8+x_9n_{\rm eff}+x_{10}n_{\rm eff}^2+x_{11}\mathcal{C} \right)\nonumber \\
 \tilde{\gamma}&=&\gamma+\mathcal{F}\left(x_{12}+x_{13}n_{\rm eff}+x_{14}n_{\rm eff}^2+x_{15}\mathcal{C} \right)\nonumber \\
 {{\rm log}_{10}\tilde{a}}&=&{\rm log}_{10}\left[a+\mathcal{F}\left(x_{16}+x_{17}n_{\rm eff}+x_{18}n_{\rm eff}^2+x_{19}\mathcal{C} \right)\right]\nonumber \\
 {{\rm log}_{10}\tilde{b}}&=&{\rm log}_{10}\left[b+\mathcal{F}\left(x_{20}+x_{21}n_{\rm eff}+x_{22}n_{\rm eff}^2+x_{23}\mathcal{C} \right)\right]\nonumber \\
 {{\rm log}_{10}\tilde{c}}&=&{\rm log}_{10}\left[c+\mathcal{F}\left(x_{24}+x_{25}n_{\rm eff}+x_{26}n_{\rm eff}^2+x_{27}\mathcal{C} \right)\right]\nonumber \\
 {{\rm log}_{10}\tilde{\mu}}&=&{\rm log}_{10}\left[\mu+ \mathcal{F}\left(x_{28}+x_{29}n_{\rm eff}+x_{30}n_{\rm eff}^2+x_{31}\mathcal{C} \right)\right]\nonumber \\
 {{\rm log}_{10}\tilde{\nu}}&=&{\rm log}_{10}\left[\nu+ \mathcal{F}\left(x_{32}+x_{33}n_{\rm eff}+x_{34}n_{\rm eff}^2+x_{35}\mathcal{C} \right)\right] \nonumber \\
\xi&=&{\rm exp}\left[\mathcal{D}\left(x_{36}+x_{37}n_{\rm eff}+x_{38}n_{\rm eff}^2+x_{39}\mathcal{C} \right)\right]\ea where $\mathcal{F}\equiv |f_{R0}|/(3\times10^{-5})$, and \be\label{eq:D}\mathcal{D}\equiv \left\vert\frac{P(k)_{\rm HS}^{\rm lin.}}{P(k)_{\Lambda \rm {CDM}}^{\rm lin.}}-{\rm max}\left[\frac{P(k)_{\rm HS}^{\hf}}{P(k)_{\Lambda {\rm CDM}}^{\hf}},1\right] \right\vert\ee 

After optimising the parameters $\vec{X}$ using the Powell's method \citep{Powell}, the coefficients are found to be,  
 
 \begin{table}[htdp]

\begin{center}
\begin{tabular}{llll}
 $x_{1}= -0.832105$;&$x_{2}=-0.238632$;&$x_{3}=0.427827$&\\
 $x_{4}=-3.367256$;&$x_{5}=3.888473$;&$x_{6}=2.294713$;&$x_{7}=8.821165$\\ 
 $x_{8}=-0.318559$;&$x_{9}=2.963588$;&$x_{10}=1.551244$;&$x_{11}=1.150983$\\ 
 $x_{12}= 2.971117$;&$x_{13}=-1.702803$;&$x_{14}=-1.284630$;&$x_{15}=-6.797889$\\ 
 $x_{16}=1.943697$;&$x_{17}=7.776061$;&$x_{18}=3.186278$;&$x_{19}=6.916149$\\    
 $x_{20}=0.999088$;&$x_{21}=8.480852$;&$x_{22}=3.644990$;&$x_{23}=9.519407$\\ 
 $x_{24}=1.934338$;&$x_{25}=2.511626$;&$x_{26}=0.792323$;&$x_{27}=0.337545$\\  
 $x_{28}=1.440371$;&$x_{29}=1.819927$;&$x_{30}=0.564780$;&$x_{31}=0.274286$\\ 
 $x_{32}=-2282.5327$;&$x_{33}=-2135.1213$;&$x_{34}=-2258.1919$;&$x_{35}=-2378.1342$\\ 
 $x_{36}=-10.656456$;&$x_{37}=-0.995708$;&$x_{38}=1.169303$;&$x_{39}=17.519593$\\  
\end{tabular}
\end{center}
\label{default}
\end{table}%

Note that in Eq (\ref{eq:D}) we require $P(k)_{\rm HS}^{\hf} \geqslant P(k)_{\Lambda {\rm CDM}}^{\hf}$ in $\mathcal{D}$ to avoid the unphysical artifact in $P(k)_{\rm HS}^{\hf}$, which is a naively application of \hf~to the HS model. But this may result in another artifact when $|f_{R0}|$ is small, namely, the resulting $P(k)$ might not be smooth. We remove this artifact by convolving the fractional difference $\Delta_P(k) \equiv P(k)_{\rm HS}^{\mgh} / P(k)_{\Lambda {\rm CDM}}^{\hf}$ with a Gaussian kernel to smooth it, \ie,

\be \tilde{\Delta}_P(k) = \frac{\int {\Delta}_P(k') e^{-\frac{({\rm ln}k-{\rm ln}k')^2}{2\sigma_k^2}} {\rm d}k'}{\int e^{-\frac{({\rm ln}k-{\rm ln}k')^2}{2\sigma_k^2}} {\rm d}k'}\ee and we find that setting the smoothing dispersion $\sigma_k$ as follows works well in practice, \ie, the smoothed $\tilde{\Delta}_P(k)$ fits to the simulation better than the unsmoothed one ${\Delta}_P(k)$.

\[ \sigma_k=\left\{ \begin{array}{ll}
         0.25\left(\frac{10^{-4}}{|f_{R0}|}\right)^{0.375} & \mbox{if $|f_{R0}| \geq 10^{-6}$};\\
        1.4 & \mbox{if $|f_{R0}| < 10^{-6}$}.\end{array} \right. \] 
 Finally $P(k)_{\rm HS}^{\mgh}$ is assembled as, \be P(k)_{\rm HS}^{\mgh}=\left[1+ \tilde{\Delta}_P(k)\right]P(k)_{\Lambda {\rm CDM}}^{\hf}\ee

\newpage

\section{Solving the HS model on linear scales using {\tt MGCAMB}}
\label{sec:apdxMGCAMB}

On linear scales, the power spectrum of the HS model can be calculated using linear perturbation theory \footnote{The HS model has been implemented in the {\tt MGCAMB} code. For details of the implementation, see  \url{http://icosmology.info/website/MGCAMB.html}}. In the conformal Newton gauge, the metric is, \be ds^2=-a^2(\tau)[(1+2\Psi)d\tau^2-(1-2\Phi)d\vec{x}^2] \ee where $\tau$ denotes the conformal time and $a$ is the scale factor normalised to $1$ at present time. In Fourier space, the modification of gravity can be parametrised by two time- and scale-dependent functions $\mu(a,k)$ and $\eta(a,k)$ satisfying \citep{BZ,MGCAMB1,pogosian} (for alternative MG parametrisations, see \eg, \citealt{MGpara1,MGpara2,MGpara3,MGpara4,MGpara5,MGpara6,MGpara7,MGpara8}),  
    \ba k^2\Psi&=&-\mu(k,a)4\pi G a^2\rho\Delta \nonumber \\
     \Phi/\Psi&=&\eta(k,a) \ea where $\Delta$ is the comoving matter density perturbation. In $\Lambda$CDM, $\mu=\eta=1$, while in the HS model, 
     \ba \mu(k,a)&=&\frac{4}{3}-\frac{(a/\lambda_c)^2}{3[k^2+(a/\lambda_c)^2]} \nonumber \\
            \eta(k,a)&=&1-\frac{2k^2}{[3(a/\lambda_c)^2+4k^2]} \ea where the comoving Compton wavelength $\lambda_c$ can be calculated as, \be \lambda_c=\left[\frac{1}{3(n+1)}\frac{\bar{R}}{|\bar{f}_{R0}|}\left(\frac{\bar{R}}{\bar{R}_0}\right)^{n+1} \right]^{1/2}\ee 
          
Feeding $\mu(k,a)$ and $\eta(k,a)$ to {\tt MGCAMB} \citep{MGCAMB1, MGCAMB2}, one can obtain the linear matter power spectrum for the HS model, shown as the dashed lines in Figs \ref{fig:planck+wmap9}, \ref{fig:wmap7},  and \ref{fig:cmass-euclid}.

\newpage
 
 \section{The PPF correspondence}

To model the nonlinear power spectrum of the HS model $P_{\rm HS}(k,z)$, the Parametrized Post-Friedmann (PPF) formula was developed by \citet{ppf}. The idea is that $P_{\rm HS}(k,z)$ is bounded between $P_{\rm nonGR}$ and $P_{\rm GR}(k,z)$, which are the nonlinear $P(k)$ in the HS model without the chameleon mechanism, and for the GR model, respectively. \citet{ppf} suggested to design a weighting function $c_{\rm nl}\Sigma^2(k,z)$ to interpolate between these two extreme cases to obtain $P_{\rm HS}(k,z)$, namely, 
\be \label{eq:PPF} P_{\rm HS}(k,z)= \frac{P_{\rm nonGR}(k,z)+c_{\rm nl}\Sigma^2(k,z)P_{\rm GR}(k,z)}{1+c_{\rm nl}\Sigma^2(k,z)} \ee
where \be \Sigma^2(k,z) = \frac{k^3}{2\pi^2} P_{\rm L}(k,z) \ee with $P_{\rm L}(k,z)$ being the linear power spectrum and $c_{\rm nl}$ is a free parameter to be fitted. $P_{\rm nonGR}(k,z)$ can be found by performing $N$-body simulations for the linearised HS model, in which the chameleon screening is effectively switched off \citep{Oyaizu2,Oyaizu3,Zhao11}.  

However, \citet{KK09} found that this setting does not fit the simulation well, and they proposed an revision for $\Sigma^2(k,z)$, namely, \be \Sigma^2(k,z) = \left[\frac{k^3}{2\pi^2} P_{\rm L}(k,z)\right]^{1/3}\ee The revised PPF formula was used to fit the simulation result in \citet{Oyaizu2,Oyaizu3} for the WMAP7 background cosmology and found to work well up to $k\sim0.5$ h/Mpc. The validity on smaller scales wasn't well tested due to the relative low resolution of the simulation performed in \citet{Oyaizu2,Oyaizu3}. 

With higher-resolution simulations performed by \citet{Zhao11}, the validity of the PPF formula was further tested on smaller scales up to $k\sim10$ h/Mpc, and it was found that PPF can capture the simulation very well for a given cosmology if $\Sigma^2(k,z)$ is further generalised to \be \label{eq:sigma} \Sigma^2(k,z) = \left[\frac{k^3}{2\pi^2} P_{\rm L}(k,z)\right]^{\alpha+\beta k^{\gamma}}\ee and the best fit values for the parameters $c_{\rm nl}, \alpha, \beta, \gamma$ were derived for a given cosmology and a given HS parameters $f_{R0}$ ($n$ is fixed to be 1) of the HS model \citep{Zhao11}.  

However, {it is difficult to} generlise the PPF approach to a fitting formula for an arbitrary $f_{R0}$ and arbitrary cosmological parameters, and the reasons include,  

\begin{description}
\item[(I)] $P_{\rm nonGR}(k,z)$ is unknown on nonlinear scales for an arbitrary cosmology, and it is hard to model it without $N$-body simulations. A naive application of Halofit does not work well (cf Fig 5 in \citealt{Zhao11});
\item[(II)] Even if we can find a fitting formula for $P_{\rm nonGR}(k,z)$ and calibrate the coefficients using simulations, it is not easy to design a fitting formula for $\Sigma^2(k,z)$ to capture the screening effect at various redshifts for various $|f_{\rm R0}|$ and $\Omega_{\rm M}$. One can see this by inverting Eq (\ref{eq:PPF}) to obtain, \be c_{\rm nl}\Sigma^2(k,z) = \frac{P_{\rm nonGR}(k,z)-P_{\rm HS}(k,z)}{P_{\rm HS}(k,z)-P_{\rm GR}(k,z)}\ee So $\Sigma$ needs to be huge to recover GR (for small $|f_{R0}|$) and vanishing when chameleon does not work (large $|f_{R0}|$). We have actually attempted to take the form of \be\label{eq:sigma2} \loge\Sigma^2(k,z) = \sum_{i=0}^5 c_i k^i\loge\left[\frac{k^3}{2\pi^2} P_{\rm L}(k,z)\right]\ee and optimised the coefficients $c_i$'s using simulations but it does not work well.  
\end{description}

Due to the above arguments, we did not take the PPF approach in this work to develop the fitting formula for $P_{\rm HS}(k,z)$.

\newpage 

\end{document}